\documentclass[prd,twocolumn,showpacs,reprint,preprintnumbers,nofootinbib,superscriptaddress,aps]{revtex4-1}

\pdfoutput=1
\usepackage{graphicx}
\usepackage{ulem}
\usepackage{listings}
\usepackage{multirow,makecell,booktabs,siunitx}
\usepackage{amsfonts,amsmath,amssymb,bm,bbm}
\usepackage{color}
\usepackage{enumitem}
\usepackage{physics}
\usepackage{url}  % for the urls in the reference.bib and then xxx.bbl
\usepackage{fancybox,framed}
\usepackage[dvipsnames, usenames]{xcolor} 
\usepackage[nolist,nohyperlinks]{acronym}
\usepackage{comment}

\newcommand{\OGW}{\Omega_\mathrm{GW}}
\newcommand{\D}{\text{d}}
\usepackage{hyperref}
\hypersetup{
    pdfnewwindow=true,      % links in new window
    colorlinks=true,       % false: boxed links; true: colored links
    linkcolor=magenta,          % color of internal links
    citecolor=magenta,        % color of links to bibliography
    filecolor=magenta,      % color of file links
    urlcolor=magenta        % color of external links
}
\usepackage{cleveref}
\def\bi{\begin{itemize}[noitemsep,leftmargin=*]
\setlength\itemsep{1em}
        }
\def\ei{\end{itemize}}

\newcommand\be{\begin{equation}}
\newcommand\ee{\end{equation}}

\newcommand{\orcid}[1]{\begingroup
  \hypersetup{hidelinks}\href{https://orcid.org/#1}{\includegraphics[width=10pt]{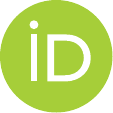}} \endgroup}
\newcommand{\del}{\partial}

\begin{document}

\title{Searching for cosmological stochastic backgrounds by notching out resolvable compact binary foregrounds with next-generation gravitational-wave detectors}

\author{Haowen Zhong \orcid{0000-0001-8324-5158}\,}
\email{zhong461@umn.edu}
\affiliation{
 School of Physics and Astronomy, University of Minnesota, Minneapolis, MN 55455, USA
}

\author{Bei Zhou \orcid{0000-0003-1600-8835}\,}
\email{beizhou@fnal.gov}
\affiliation{Theoretical Physics Department, Fermi National Accelerator Laboratory, Batavia, Illinois 60510, USA}
\affiliation{Kavli Institute for Cosmological Physics, University of Chicago, Chicago, Illinois 60637, USA}

\author{Luca Reali \orcid{0000-0002-8143-6767}\,}
\email{lreali1@jhu.edu}
\affiliation{William H. Miller III Department of Physics and Astronomy, Johns Hopkins University, Baltimore, Maryland 21218, USA
}

\author{Emanuele Berti \orcid{0000-0003-0751-5130}\,}
\email{berti@jhu.edu}
\affiliation{William H. Miller III Department of Physics and Astronomy, Johns Hopkins University, Baltimore, Maryland 21218, USA
}

\author{Vuk Mandic \orcid{0000-0001-6333-8621}\,}
\email{vuk@umn.edu}
\affiliation{
School of Physics and Astronomy, University of Minnesota, Minneapolis, MN 55455, USA
}

\date{\today}
\preprint{FERMILAB-PUB-24-0271-T}

\begin{abstract}
Stochastic gravitational-wave backgrounds can be of either cosmological or astrophysical origin. The detection of an astrophysical stochastic gravitational-wave background with ground-based interferometers is expected in the near future. Perhaps even more excitingly, the detection of stochastic backgrounds of cosmological origin by future ground-based interferometers could reveal invaluable information about the early Universe. From this perspective, the astrophysical background is a {\it foreground} that can prevent the extraction of this information from the data. In this paper, we revisit a time-frequency domain notching procedure previously proposed to remove the astrophysical foreground in the context of next-generation ground-based detectors, but we consider the more realistic scenario where we remove individually detectable signals by taking into account the uncertainty in the estimation of their parameters. We find that time-frequency domain masks can still efficiently remove the astrophysical foreground and suppress it to about $5\%$  of its original level. Further removal of the foreground formed by unresolvable events (in particular, unresolvable binary neutron stars), which is about $10$ times larger than the residual foreground from realistic notching, would require detector sensitivity improvements. Therefore, the main limitation in the search for a cosmological background is the unresolvable foreground itself, and not the residual of the notching procedure. 
%\Bei{Add one sentence in the beginning about why detecting subdominant/cosmological \ac{SGWB}? I.e., the payoffs.}
\end{abstract}
\maketitle

\begin{acronym}
    \acro{GW}{gravitational-wave}
    \acro{PSD}{power spectral density}
    \acro{GR}{general relativity}
    \acro{CBC}{compact binary coalescence}
    \acro{BH}{black hole}
    \acro{BBH}{binary black hole}
    \acro{BNS}{binary neutron star}
    \acro{NSBH}{neutron star-black hole}
    \acro{SFR}{star formation rate}
    \acro{SNR}{signal-to-noise ratio}
    \acro{LVK}{LIGO-Virgo-KAGRA}
    \acro{ET}{Einstein Telescope}
    \acro{CE}{Cosmic Explorer}
    \acro{PE}{parameter estimation}
    \acro{SGWB}{stochastic gravitational wave background}
    \acro{XG}{next-generation}
\end{acronym}

%%%%%%%%%%%%%%%%%%%%%%%%%%%%%%%%%%%
%%%%%%       SECTION        %%%%%%%
%%%%%%%%%%%%%%%%%%%%%%%%%%%%%%%%%%%

\section{Introduction}
\label{sec_intro}

% {\Bei{Comments on the whole draft:

% +) I suggest changing the terminology throughout the paper: ``best masks'' $\rightarrow$ ``ideal masks'' and ``biased masks'' $\rightarrow$ ``realistic masks''. Otherwise, the readers might think that we didn't do the best we could do.  % Haowen already made the revision.

% +) Maybe in the main text we don't need to explain which curve in which color/style is for what. The caption or legend does the job. So no need to repeat.
% }
% }

As ground-based \ac{GW} observatories %LIGO~\cite{Harry:2010zz, LIGOScientific:2014pky}, Virgo~\cite{VIRGO:2014yos} and KAGRA~\cite{Aso:2013eba} 
improve in sensitivity, a detection of the astrophysical \ac{SGWB} from \ac{CBC} events is expected in the near future~\cite{KAGRA:2021duu,KAGRA:2021kbb,Renzini:2022alw}.
%it is quite probable that we will be able to start detecting astrophysically originated \ac{SGWB} in the near future~\cite{Renzini:2022alw,KAGRA:2021duu}. 
%Such background is expected to dominate over other interesting \acp{SGWB} that might be present, e.g. those from supernova explosions~\cite{Ferrari:1998ut, Buonanno:2004tp, Crocker:2015taa, Crocker:2017agi, Finkel:2021zgf}, standard inflation~\cite{Grishchuk:1974ny,Starobinsky:1979ty,Grishchuk:1993te}, axion inflation~\cite{Barnaby:2011qe}, cosmic strings~\cite{Damour:2004kw, Siemens:2006yp, Olmez:2010bi, Regimbau:2011bm}. The task of detecting these subdominant backgrounds is thus quite challenging with current interferometers.
Besides the \ac{SGWB} from \acp{CBC}, several other interesting \acp{SGWB} might be present in the data, e.g. those from supernova explosions~\cite{Ferrari:1998ut, Buonanno:2004tp, Crocker:2015taa, Crocker:2017agi, Finkel:2021zgf}, standard inflation~\cite{Grishchuk:1974ny, Starobinsky:1979ty, Grishchuk:1993te}, axion inflation~\cite{Barnaby:2011qe}, or cosmic strings~\cite{Damour:2004kw, Siemens:2006yp, Olmez:2010bi, Regimbau:2011bm}. Detecting these subdominant \acp{SGWB} has significant scientific benefits: the subdominant astrophysical \acp{SGWB} potentially contain key information about the properties of their corresponding sources~\cite{LIGOScientific:2020kqk, KAGRA:2021duu, Bavera:2021wmw}, while cosmological \acp{SGWB} would open up a unique window to observe the earliest moments of the Universe and to probe physics at energies close to the Planck scale~\cite{Grishchuk:1974ny, Starobinsky:1979ty, Grishchuk:1993te, Barnaby:2011qe, Damour:2004kw, Siemens:2006yp, Olmez:2010bi, Regimbau:2011bm}. However, the predicted energy densities of these different \acp{SGWB} vary by many orders of magnitude, and it is challenging to resolve them with current interferometers unless their amplitude is comparable to, or larger than, the \ac{SGWB} from \acp{CBC}, which in this context is a {\em foreground} limiting our ability to observe other \acp{SGWB}.

Planned \ac{XG} ground-based detectors, such as \ac{CE}~\cite{CE} and the \ac{ET}~\cite{ET}, will see an unprecedented increase in sensitivity both to individual \ac{CBC} events and to stochastic searches, allowing us to probe \acp{SGWB} across several orders of magnitude in amplitude~\cite{Renzini:2022alw}. Moreover, they are expected to individually resolve nearly all of the \ac{BBH} events and a significant fraction of \ac{BNS} and \ac{NSBH} events across the whole Universe~\cite{CEHorizon, ET_PSD}. This opens up the possibility of removing the \ac{CBC} foreground from the data to search for other subdominant \acp{SGWB} of both cosmological and astrophysical origin~\cite{Regimbau:2016ike, Sachdev:2020bkk}.
%With the \ac{XG} of ground-based detectors like \ac{CE}~\cite{CE} and \ac{ET}~\cite{ET}, we may be able to individually resolve all of the \ac{BBH} events and and a significant fraction of \ac{BNS} and \ac{NSBH} events across the whole Universe~\cite{CEHorizon,ET_PSD}. 
%However, this unprecedented high sensitivity also results in a novel challenge: astrophysically originated \ac{SGWB} itself becomes to a new source of contamination for the detection of the relatively weaker cosmologically originated \ac{SGWB}.~\Luca{It's not due to the sensitivity. The astro \ac{SGWB} is intrinsically higher at these frequencies, it dominates over the cosmological in LIGO as well.} 

In practice, however, \ac{PE} uncertainties on the individually resolved signals imply that their subtraction is never perfect, resulting in a residual foreground that is potentially stronger than other targeted \ac{SGWB}~\cite{Zhou:2022nmt, Zhou:2022otw, Song:2024pnk}. Several techniques have been proposed to reduce this residual foreground, from applying a projection procedure in the time or frequency domain~\cite{Cutler:2005qq, Sharma:2020btq} to estimating and subtracting the expected residual power~\cite{Pan:2023naq, Li:2024iua}. Alternative techniques that simultaneously fit all the events within a Bayesian framework have also been proposed~\cite{Smith:2017vfk,Smith:2020lkj,Biscoveanu:2020gds}, although their computational feasibility in the context of \ac{XG} detectors has yet to be proven.
 
%There are multiple possible solutions have been proposed including subtracting resolvable \ac{CBC} events in either time or frequency domain~\cite{Regimbau:2016ike,subtractionSurabhi,Zhou:2022otw,Song:2024pnk}, and simultaneously estimating \ac{SGWB} with all individually detected \ac{CBC} events~\cite{Smith:2017vfk,TBS2}.~\Luca{Note to self: expand a bit here. Mention Pieroni, Arianna, etc} However, due to the uncertainty associated with \ac{PE}, solely subtracting \ac{CBC} signals out will result in a residual foreground which still might be much stronger than the targeting cosmological \ac{SGWB}. To suppress this residual foreground, two different approaches have been proposed. We can either apply a projection procedure in time or frequency domain~\cite{Cutler:2005qq,Sharma:2020btq} or estimating and subtracting the expected residual foreground~\cite{Pan:2023naq,Li:2024iua} to get cleaned energy spectrum of \ac{SGWB}.

In previous work, we proposed a novel notching procedure in the time-frequency domain (different from other approaches that work either in the time or in the frequency domain), where pixels containing \ac{CBC} signals in time-frequency space are notched out~\cite{Zhong:2022ylh}. In that previous work we assumed that all
%ac{CBC} signals are resolvable and that we can recover the 
\ac{CBC} parameters can be recovered perfectly (i.e., without uncertainty). We computed the resulting sensitivity of the stochastic search after removing all \ac{CBC} events, and calculated the strength of the \ac{SGWB} generated by the subthreshold \ac{CBC} events.
% \footnote{Even though we assumed that we could resolve all the \ac{CBC} events in the previous work~\cite{Zhong:2022ylh}, this assumption is unrealistic It was adopted to test whether the notching procedure would remove too many pixels in the time-frequency domain, potentially resulting in very poor sensitivity for the further detection of cosmological \ac{SGWB}.} 
Our results showed that this unresolvable \ac{CBC} foreground was much stronger than the detector noise after foreground cleaning. This suggested that the notching procedure can effectively remove the astrophysical foreground, and that the primary limitation for future cosmological \ac{SGWB} searches will not be the detector sensitivity, but rather the limit set by the unresolvable \ac{CBC} foreground. 

In this work, we consider a more realistic scenario where only resolvable events can be notched out, and the realistic recovery of \ac{CBC} parameters is taken into account. We generate \ac{BNS} and \ac{BBH} populations consistent with the \ac{LVK} catalog~\cite{KAGRA:2021duu}. We adopt a Fisher-matrix approach~\cite{Finn:1992wt} to estimate the errors on the inferred parameters for resolved events. We then extend our previous procedure to devise a mask for the recovered signals in the time-frequency plane. \textit{We find that even when only resolvable events are removed and the masks used for notching are determined by the recovered \ac{CBC} parameters, we are still able to remove the foreground from resolved events and suppress the whole foreground generated by all \ac{CBC} events, including the unresolvable ones, to $\sim 5.4\%$ of its original level.} The sensitivity to other subdominant \acp{SGWB} is thus limited by the foreground from unresolved \ac{BNS} events, with the residuals from realistic subtraction (of resolved events) providing an additional $\sim 10\%$ contribution in the remaining foreground after notching.

The rest of the paper is structured as follows. In Sec.~\ref{sec_sgwb} we review the standard process for cross-correlation searches of \acp{SGWB}. In Sec.~\ref{sec_params} we provide the details of our \ac{CBC} population models. In Sec.~\ref{sec_fisher} we describe the Fisher-matrix formalism used to estimate the uncertainties associated with \acp{CBC}, along with our choices of detector network and waveform model. In Sec.~\ref{sec_mockdata} we discuss our procedure to generate mock data and compute masks for the subsequent notching. In Sec.~\ref{sec_notching} we show the notching results. In Sec.~\ref{sec_dis} we draw our conclusions and outline directions for future work.
In Appendix~\ref{app_powerlaw} we discuss the validity of the power-law approximation for the \ac{SGWB} from \acp{CBC} commonly used in the literature.
% \eb{Appendices}
% \bi
% \item Why detect \ac{SGWB}? Because \ac{SGWB} is important...

% \item Ground-based detectors...  

% \item Foreground... 

% \item Different methods. Subtraction, notching, global fitting, etc.

% \item In this work, we ...

% \item Layout ...

% \ei

%%%%%%%%%%%%%%%%%%%%%%%%%%%%%%%%%%%
%%%%%%       SECTION        %%%%%%%
%%%%%%%%%%%%%%%%%%%%%%%%%%%%%%%%%%%
\section{Cross-correlation search for the \ac{SGWB}}\label{sec_sgwb}
The dimensionless \ac{SGWB} energy density spectrum is defined by
\begin{equation}
    \OGW(f):=\frac{f}{\rho_{c,0}}\frac{\D \rho_\mathrm{GW}(f)}{\D f},
\end{equation}
where $\D\rho_\mathrm{GW}(f)$ is the \ac{GW} energy density in the frequency bin $(f,f+\D f)$, $\rho_{c,0}=3H_0^2c^2/(8\pi G)$ is the critical energy density to close the Universe, $c$ is the speed of light, $G$ is Newton's constant, and $H_0=67.66$ km/s/Mpc~\cite{Planck:2018vyg} is the Hubble constant.

In the \ac{SGWB} search, we assume a power-law \ac{SGWB} of the form
\begin{equation}
    \OGW(f)=\Omega_\alpha \left( \frac{f}{f_\mathrm{ref}} \right)^\alpha,
\end{equation}
where $\alpha$ is the power-law index of the \ac{GW} spectrum, and we choose $f_\mathrm{ref}=25$\,Hz as the reference frequency.

For terrestrial \ac{GW} detectors, we need to cross-correlate the data from a network consisting of two or more detectors to pick up the common \ac{SGWB} signal. For a single \ac{GW} detector, we can write the strain time series as
\begin{equation}
    h(t)=F^+h_+(t)+F^\times h_\times(t),
\end{equation}
where $F^{+/\times}$ is the antenna pattern function corresponding to the $+$ and $\times$ \ac{GW} polarization modes, respectively. In a real \ac{SGWB} search, we usually cut the time strain data into chunks with duration $T=60-192$\,s~\cite{KAGRA:2021kbb, Renzini:2022alw}. However, in our study, we are required to capture the transient features of \ac{CBC} tracks in the time-frequency domain, which calls for a proper time and frequency resolution. Therefore we choose $T=4$\,s, as in Ref.~\cite{Zhong:2022ylh}. We denote the Fourier transform of these segments by $\tilde{h}_\mathrm{I}(t_i;f_j)$, where the index ``I'' refers to the Ith detector in the network, $t_i$ is the time segment used for the analysis, and $f_j$ indicates the frequency bin. We define the standard cross-correlation statistics and the corresponding variance estimator for the baseline IJ as follows~\cite{KAGRA:2021kbb, Allen:1997ad}:
\begin{eqnarray}
    \hat{C}_{\text{IJ}}(t_i;f_j) & = & \Big( \frac{20\pi^2f_j^3}{3H_0^2T} \Big) \frac{\Re[{\tilde{h}_{\rm{I}}^*(t_i;f_j)\tilde{h}_{\rm{J}}(t_i;f_j)}]} {\gamma_{\text{IJ}}(f_j)} \nonumber \\
    \hat{\sigma}_{\text{IJ}}^2(t_i;f_j) & = & \Big( \frac{20\pi^2f_j^3}{3H_0^2} \Big)^2 \frac{P_{n_{\rm{I}}}(t_i;f_j)P_{n_{\rm{J}}}(t_i;f_j)} {8T\Delta f\gamma_{\rm{IJ}}^2(f_j)}.
    \label{eq:stoch}
\end{eqnarray}
Here $\gamma_{\mathrm{IJ}}(f_j)$ is the overlap reduction function between the Ith and Jth detectors~\cite{Allen:1997ad}, $\Delta f=0.25$\,Hz is the Fourier transform resolution, and $P_{n\mathrm{I}}(t_i;f_j)$ is the \ac{PSD} of the Ith detector at time $t_i$. 

As some of us pointed out in previous work~\cite{Zhong:2022ylh}, directly estimating the detector noise \ac{PSD} from the data is particularly challenging, since the presence of the \ac{CBC} foreground could lead to large biases. Here we assume that the detector noise is perfectly modeled, and we adopt the true PSD used for generating detector noise data to carry out the above calculations. 

%could be largely biased due to the existence of the astrophysical foreground. Hence, instead of estimating PSD using real data, we assume that we fully understand the detector noise and take the true PSD that is used for generating the detector noise data to carry out the above calculations. Again, we recognize estimating PSD without bias with real data as another big challenge in the future, since we have no way to know the true PSD beforehand as \textit{a priori}.~\Luca{Rephrase a bit}

Having $\hat{C}_{\mathrm{IJ}}(t_i;f_j)$ and $\hat{\sigma}^{2}_{\mathrm{IJ}}(t_i;f_j)$ at hand, we can combine them to maximize the resulting signal-to-noise ratio:
\begin{eqnarray}
\hat{C}_{\text{IJ}} & = & \frac{\displaystyle{\sum_{ij} w(f_j) \hat{C}_{\text{IJ}}(t_i;f_j) \hat{\sigma}_{\text{IJ}}^{-2}(t_i;f_j)}} {\displaystyle{\sum_{ij} w^2(f_j) \hat{\sigma}_{\text{IJ}}^{-2}(t_i;f_j)}},\nonumber\\
\hat{\sigma}_{\text{IJ}}^{-2} & = & \sum_{ij} \frac{w^2(f_j)}{\hat{\sigma}^2_{\text{IJ}}(t_i;f_j)},
\label{eq:combine}
\end{eqnarray}
with weights defined by $w(f_j)=\Omega_{\text{GW}}(f_j)/\Omega_{\alpha}$~\cite{KAGRA:2021kbb}. With these definitions, $\hat{C}_{\text{\text{IJ}}}$ is the broadband detection statistic averaged over the entire observation data, normalized so that $\langle \hat{C}_{\text{\text{IJ}}} \rangle = \Omega_{\rm \alpha}$. We can perform a weighted average over all time bins of $\hat{C}_\mathrm{IJ}(t_i;f_j)$ and $\hat{\sigma}^{-2}_\mathrm{IJ}(t_i;f_j)$ to get two frequency-domain spectra, $\hat{C}_\mathrm{IJ}(f)$ and $\sigma_\mathrm{IJ}^{-2}(f)$, of the \ac{SGWB} and detector noise:
\begin{eqnarray}
\hat{C}_{\text{IJ}}(f_j) & = & \frac{\displaystyle{\sum_{i} \hat{C}_{\text{IJ}}(t_i;f_j) \hat{\sigma}_{\text{IJ}}^{-2}(t_i;f_j)}} {\displaystyle{\sum_{i} \hat{\sigma}_{\text{IJ}}^{-2}(t_i;f_j)}},\nonumber\\
\hat{\sigma}_{\text{IJ}}^{-2}(f_j) & = & \sum_{i} \frac{1}{\hat{\sigma}^2_{\text{IJ}}(t_i;f_j)}.
\label{eq:combine_t}
\end{eqnarray}

If we have more than two detectors in the network, we can further combine $\hat{C}_{\mathrm{IJ}}$ and $\hat{\sigma}_{\mathrm{IJ}}^{-2}$ among different pairs to obtain
\begin{eqnarray}
\hat{C}&=&\frac{\displaystyle{\sum_\mathrm{IJ}\hat{C}_{\mathrm{IJ}}\hat{\sigma}_{\mathrm{IJ}}^{-2}}}{\displaystyle{\sum_{\mathrm{IJ}}\hat{\sigma}_\mathrm{IJ}^{-2}}},\nonumber\\
\hat{\sigma}^{-2}&=&\sum_\mathrm{IJ}\frac{1}{\hat{\sigma}_\mathrm{IJ}^2},
\label{eq:baselinecombine}
\end{eqnarray}
where IJ denotes the baseline composed by the Ith and Jth detectors, and the summation goes through all possible baselines. Again, we can get spectra $\hat{C}(f)$ and $\hat{\sigma}^{-2}(f)$ by a weighted average of $\hat{C}_\mathrm{IJ}(f)$ and $\hat{\sigma}_\mathrm{IJ}(f)$:
\begin{eqnarray}
\hat{C}(f)&=&\frac{\displaystyle{\sum_\mathrm{IJ}\hat{C}_{\mathrm{IJ}}(f)\hat{\sigma}_{\mathrm{IJ}}^{-2}(f)}}{\displaystyle{\sum_{\mathrm{IJ}}\hat{\sigma}_\mathrm{IJ}^{-2}(f)}},\nonumber\\
\hat{\sigma}^{-2}(f)&=&\sum_\mathrm{IJ}\frac{1}{\hat{\sigma}_\mathrm{IJ}^2(f)}.
\label{eq:baselinecombine}
\end{eqnarray}

%%%%%%%%%%%%%%%%%%%%%%%%%%%%%%%%%%%
%%%%%%       SECTION        %%%%%%%
%%%%%%%%%%%%%%%%%%%%%%%%%%%%%%%%%%%
\section{Compact Binary Populations}
\label{sec_params}

We use the same \ac{CBC} population models as in Ref.~\cite{Zhong:2022ylh}, with the mass and redshift distributions described in Secs.~\ref{sec_mass_dist} and \ref{sec_redshift_dist}, respectively. In the simulations, we consider a one-year observation time.

\subsection{Mass and angle distributions}
\label{sec_mass_dist}

For the \ac{BBH} mass distribution, we employ the \texttt{PowerLaw+Peak} model from the latest \ac{LVK} catalog~\cite{KAGRA:2021duu}. For \ac{BNS} events, we adopt a uniform mass distribution ranging from $1M_\odot$ to $2M_\odot$, in accordance with~\cite{KAGRA:2021duu, Landry:2021hvl}. We assume the spins to be zero for all binary components, as we expect their impact to be 
subdominant for our notching procedure. Recent work pointed out that the inclusion of spin effects could lead to 10 times worse and 20 times worse results for \acp{BBH} and \acp{BNS}, respectively, when performing subtraction~\cite{Song:2024pnk}. However, the masks we use for notching are primarily determined by the detector-frame chirp masses, whose inference is not significantly affected by the inclusion of parameters that appear at higher order in the expansion of the phase, such as the spins. For the same reason, we also
%For analogous reasons, we 
neglect the effects of nonzero tidal deformability for \acp{BNS}. 
The angular parameters for both \ac{BBH} and \ac{BNS} binaries (i.e., inclination angle $\iota$, right ascension $\alpha$, declination $\delta$, and polarization angle $\psi$) are drawn isotropically. We sample the coalescence phase $\phi_{\rm c}$ uniformly within $\left[0,2\pi\right]$.

%\Luca{How long is the observation time we consider, or equivalently how many events do we simulate in total?}

%The angular parameters for both \ac{BBH} and \ac{BNS} models—including the cosine of the inclination angle, sky localization angles, and \ac{GW} polarization—are sampled isotropically. %based on a uniform distribution. Furthermore, we proceed under the assumption that both \ac{CBC} components are non-spinning.~\Luca{Neglect tidal deformability, eccentricity...}

%%%%%%%%%%%%%%%%%%%%%%%%%%%%%%%%%%%%%%%%%%%%%%%%%%
\subsection{Redshift distribution and merger rate}
\label{sec_redshift_dist}

As in Refs.~\cite{Zhong:2022ylh,Zhou:2022nmt,Zhou:2022otw}, we draw the redshift $z$ from a distribution $p(z)$ such that
\begin{equation}
    p(z)=\frac{R_z(z)}{\displaystyle{\int_0^{10}}R_z(z)\D z},
\end{equation}
where $R_z(z)$ is the \ac{CBC} merger rate per redshift interval over the range $z\in[0,10]$ in the detector frame:
\begin{equation}
    R_z(z)=\frac{R_m(z)}{1+z}\frac{\D V}{\D z}\Big|_z,
\end{equation}
$\D V/\D z$ is the comoving volume element, and $R_m(z)$ is the merger rate per comoving volume in the source frame:
\begin{equation}
    R_m(z)=\int_{t_{\min}}^{t_{\max}}R_f(z_f)P(t_d)\D t_d.
\end{equation}
Here, $R_f(z_f)$ is the binary formation rate as a function of the redshift at binary formation time $t_f=t_f(z_f)$, and $P(t_d)$ is the distribution of the time delay $t_d$ between binary formation and merger. For \acp{BNS}, we set $t_{\min}=20$\,Myr, and for \acp{BBH}, we set $t_{\min}=50$\,Myr; $t_{\max}$ is set to be the Hubble time in both cases. We use a time delay distribution $p(t_d) \propto 1/t_d$. We neglect the time needed for two stars to form a binary system and simply model the binary formation rate as following the \ac{SFR}~\cite{Finkel:2021zgf}:
\begin{equation}
    \text{SFR}(z)=\nu\frac{pe^{q(z-z_m)}}{p-q+qe^{p(z-z_m)}},
\end{equation}
with %$\nu=0.178M_\odot/\text{yr}/\text{Mpc}^3,
$z_m=2.00$, $p=2.37$ and $q=1.80$.

We set the redshift range to $z\in[0,10]$, and we choose the normalization factors $\nu_{\text{BBH}}$ and $\nu_{\text{BNS}}$ to be consistent with the inferred \ac{LVK} merger rates~\cite{KAGRA:2021duu}.
%so as to match the \ac{BBH} and \ac{BNS} merger rates to the observed ones~\cite{Population}. 
Specifically, we fix $\nu_{\text{BBH}}$ so that $R_{\rm BBH}(z=0.2) = 28.1 {\rm \; Gpc^{-3} \; yr^{-1}}$, as in Ref.~\cite{Zhong:2022ylh}. For \acp{BNS}, since performing the analysis for all of the three populations considered in Ref.~\cite{Zhong:2022ylh} is computationally expensive, we consider only the MS Model~\cite{KAGRA:2021duu}, fixing $R_{\rm BNS} = 470 {\rm \; Gpc^{-3} \; yr^{-1}}$. Both of the rates listed above are well within the uncertainty bands assumed in Refs.~\cite{Zhou:2022nmt,Zhou:2022otw}.

% \item Model: 
%     \begin{itemize}
%         \item Masses: \ac{BBH} power law+peak from GWTC-3, \ac{BNS} uniform distribution.
%         \item Redshift: Star formation rate + time delay (and metallicity cut for \ac{BBH}s?)
%         \item Rates: from latest LVK catalog (GWTC-3). In Haowen's paper~\cite{Zhong:2022ylh}, \ac{BNS} rates are a bit lower compared to Bei's and Luca's~\cite{Zhou:2022nmt,Zhou:2022otw} because of the choice of mass distribution, which is the same that we are adopting here. Should we stick to this choice or try using the highest rates, at least for the most pessimistic case?\Haowen{We can wait for the results of the medium rate case and run the pessimistic case later. To generate time series and also the cross-correlation results for 5 detectors needs huge amount of storage ($\mathcal{O}(10)$ TB for each population).}
%     \end{itemize}

% \item Detector network:  network\_spec = ['CE-40\_H','CE-40\_L','ET\_ET1','ET\_ET2','ET\_ET3'] ?

% ['CE-40\_CEA','CE-40\_CEB','ET\_ET1','ET\_ET2','ET\_ET3']

% \item Frequency range 3--2048 Hz

% \item WF model: IMRPhenomXAS

% \item Earth rotation: not include

%%%%%%%%%%%%%%%%%%%%%%%%%%%%%%%%%%%
%%%%%%       SECTION        %%%%%%%
%%%%%%%%%%%%%%%%%%%%%%%%%%%%%%%%%%%

%%%%%%%%%%%%%%%%%%%%%%%%%%%%%%%%%%%
%%%%%%       SECTION        %%%%%%%
%%%%%%%%%%%%%%%%%%%%%%%%%%%%%%%%%%%
\section{Errors from realistic recovery}
\label{sec_fisher}

We now discuss the Fisher matrix formalism we use to estimate the errors on the inferred \ac{CBC} parameters, along with technical details on the detector network and waveform models chosen for our analysis. 
%The consideration and procedures are similar to our previous work~\cite{Zhou:2022nmt, Zhou:2022otw}. 

We consider a network of three \ac{XG} detectors: two 40-km-long \ac{CE} located at LIGO Hanford and LIGO Livingston~\cite{Harry:2010zz,LIGOScientific:2014pky}, and one \ac{ET} at the location of Virgo~\cite{VIRGO:2014yos}. This extends our previous work, where we focused on CE only~\cite{Zhong:2022ylh}. For the \ac{CE} detectors, we assume the noise \ac{PSD} corresponding to the second stage of development~\cite{CE2_PSD}, while for \ac{ET}, we assume the \ac{PSD} for the triangular configuration~\cite{ET_PSD}. When we discuss detector networks, we denote the detectors located at Hanford and Livingston as H and L, for brevity. Since one ET is composed of three colocated detectors, we denote them by ET$i~(i=1,2,3)$. 

For both \acp{BBH} and \acp{BNS}, we use the \texttt{IMRPhenomXAS} waveform model~\cite{Pratten:2020fqn}, which is quasi-circular and includes only the dominant mode of the radiation. We neglect Earth-rotation effects, which were found to have a small impact on our results~\cite{Zhou:2022nmt, Zhou:2022otw}. We set the low-frequency sensitivity limit to be $5\,\rm{Hz}$.

In order to assess whether a certain \ac{CBC} event is detectable, we calculate the matched-filtering \ac{SNR}:
%of all \ac{CBC} events in our list with {\tt GWBENCH}. Given a single detecotr, and a \ac{CBC} event whose frequency domain signal can be denoted as $h(f)$, the \ac{SNR} is defined by:
\begin{equation}
    \mathrm{SNR}=\sqrt{\langle h(f)|h(f)\rangle}.
    \label{eq_SNR}
\end{equation}
Here $\langle\cdot|\cdot\rangle$ is the usual signal inner product,
\begin{equation}
\langle a|b\rangle = 4\,\mathrm{Re} \int_0^{\infty}\frac{\tilde{a}(f)\tilde{b}^*(f)}{P_n(f)}\,\D f \,,
\label{eq_fisher}
\end{equation}
where $a$ and $b$ are two generic signals, and we denote the complex conjugate with an asterisk.

Due to the large size of our catalog, performing a full Bayesian \ac{PE} for each resolved event is too computationally expensive. Therefore, we estimate the errors within the linear signal approximation~\cite{Finn:1992wt}. The linear signal approximation is strictly valid only in the high-SNR limit. However, it provides the (Cramer-Rao) lower bound on the variance of the maximum likelihood estimator of the parameters, if it is unbiased~\cite{Vallisneri:2007ev}. In this sense, our results are likely to provide an optimistic estimate of the parameter errors. We model the posterior probabilities for the parameters of each source as a multivariate normal distribution. The covariance matrix is given by the inverse of the Fisher matrix~\cite{Finn:1992wt}
%The center of the distribution is at the true values of the parameters and the covariance matrix is given by the inverse of the information matrix, $\Gamma$, which is defined as
\begin{equation}
\Gamma_{\alpha\beta} = \left\langle\frac{\del h}{\del\theta^\alpha} \middle| \frac{\del h}{\del\theta^\beta}\right\rangle \,,
\end{equation}
%for a single \ac{GW} detector.  $P_n(f)$ is the single sided detector noise \ac{PSD}.
with $\theta^\alpha$ denoting the event parameters. As in Refs.~\cite{Zhou:2022nmt,Zhou:2022otw}, we consider a set of $9$ parameters:
\begin{equation}
\bm{\theta}=
\left\{ 
\ln\left(\frac{\mathcal{M}_z}{M_\odot}\right), 
\eta, 
\ln\left(\frac{D_L}{\text{Mpc}}\right), 
\cos\iota, 
\cos\delta,
\alpha, 
\psi,
\phi_{c}, 
t_{c} 
\right\} \,.
\label{eq_info_paras}
\end{equation}
Here, $\mathcal{M}_c^z = \mathcal{M}_c (1+z)$ is the detector-frame chirp mass, with $\mathcal{M}_c=(m_1m_2)^{3/5}/(m_1+m_2)^{1/5}$, and $m_{1,2}$ the component masses; $D_{\rm L}$ is the luminosity distance, obtained from the sampled redshift assuming the Planck 2018 cosmology~\cite{Planck:2018vyg}; and $\eta=(m_1m_2)/(m_1+m_2)^2$ is the symmetric mass ratio.
%Source frame chirp mass is defined to be $\mathcal{M}_c=(m_1m_2)^{3/5}/(m_1+m_2)^{1/5}$. We ignore the 6 spin parameters because their effects are expected to be subdominant~\cite{Zhou:2022nmt, Zhou:2022otw}.~\Luca{Repetition}
Both Eqs.~\eqref{eq_SNR}, \eqref{eq_fisher} can be naturally extended to a network of $N_{\rm det}$ detectors as
%With a network of detectors consists of $N_\mathrm{det}$ detectors, the network \ac{SNR} reads:
\begin{eqnarray}
\mathrm{SNR}_\mathrm{net} &=& \sqrt{\sum_{i=1}^{N_\mathrm{det}}\mathrm{SNR}_i}\,, \nonumber \\
\Gamma &=& \sum_{j=1}^{N_{\rm det}} \Gamma_j\,,
\label{eqs_detenet}
\end{eqnarray}
where we assume that the noise is uncorrelated among different detectors. To compute the Fisher matrices, we use the public \texttt{python} package \texttt{GWBENCH}~\cite{Borhanian:2020ypi}.
%where summation goes through all detectors in the network.
%For a network consisting of $N_{\rm det}$ detectors with independent noise characteristics, the combined information matrix, $\Gamma$, is computed as the sum of the information matrices from each individual detector:
%
%\begin{equation}
%\Gamma = \sum_{j=1}^{N_{\rm det}} \Gamma_j.
%\label{eq_infonet}
%\end{equation}
%We use {\tt GWBENCH}~\cite{Borhanian:2020ypi} to calculate the information matrices. We do not include the Earth-rotation effects, which were found to be small for parameter estimation~\cite{Zhou:2022nmt, Zhou:2022otw}. 

We neglect the impact of overlapping signals for both detection and \ac{PE}, estimating the errors on the inferred parameters of each signal independently. For \ac{XG} detectors, signal confusion has been shown to slightly decrease the detectability of the \ac{CBC} events at high redshift~\cite{Wu:2022pyg}. The impact on \ac{PE} is expected to be mild~\cite{Reali:2022aps,Reali:2023eug,Johnson:2024foj} unless two or more signals merge within $\sim 0.5\,\rm{s}$ from each other~\cite{Himemoto:2021ukb, Pizzati:2021apa, Samajdar:2021egv, Janquart:2022nyz}.
%First of all, we ignore the overlapping between many signals and do the parameter estimation for each signal independently. The overlapping could slightly decrease the detectability of the \ac{CBC} events~\cite{Wu:2022pyg}. Moreover, the parameter estimation is more complicated when fitting multiple overlapping signals simultaneously~\cite{Himemoto:2021ukb, Pizzati:2021apa, Samajdar:2021egv, Janquart:2022nyz}. Additionally, the presence of weak overlapping signals induces a ``confusion noise'' component on top of the instrument noise, hence increasing the estimated parameter uncertainties~\cite{Antonelli:2021vwg, Reali:2022aps}. However, these effects would not affect our results qualitatively.

%%%%%%%%%%%%%%%%%%%%%%%%%%%%%%%%%%%
%%%%%%       SECTION        %%%%%%%
%%%%%%%%%%%%%%%%%%%%%%%%%%%%%%%%%%%
\section{Mock Data Generation and Computation of the Masks}\label{sec_mockdata}

In this section, we introduce the way we generate mock strain data and how we determine masks to notch out the astrophysical foreground.

We use the \texttt{MDC\_Generation} package~\cite{Meacher:2015iua, MDC} to generate a list of \ac{CBC} parameters over a one-year long period, based on the population models outlined in Sec.~\ref{sec_params} We then use \texttt{Bilby}~\cite{Ashton:2018jfp} to generate time-domain gravitational waveforms %for \ac{CBC} events using 
with \texttt{IMRphenomXAS} and simulate the strain noise series based on the expected noise \ac{PSD}. We choose the minimum injection frequency to be $5\,\rm{Hz}$. 
The full time strain series %containing \ac{CBC} signals and noise 
consists of the superposition of the pure \ac{CBC} injections and the noise strain series generated separately in this way.

Compared to our previous work~\cite{Zhong:2022ylh}, where we assumed that we could recover all \ac{CBC} parameters perfectly, here we consider a more realistic scenario in which only resolvable signals can be removed and there is a \ac{PE} uncertainty associated with the recovered \ac{CBC} parameters.

For \acp{BBH}, we use a detection threshold $\rho^\mathrm{BBH}_\mathrm{thr}=8$. 
For \acp{BNS}, we set $\rho^\mathrm{BNS}_\mathrm{thr}=12$ to reduce the chance of false detections. This is because \acp{BNS}, having smaller masses compared to \acp{BBH}, generally emit signals with weaker amplitudes during the inspiral phase, making it more challenging to distinguish them from detector noise~\cite{Zhou:2022nmt, Zhou:2022otw, Sachdev:2020lfd, Regimbau:2016ike}. %Since applying real \ac{PE} is quite computationally expensive, we approximate the uncertainty in \ac{PE} by Fisher matrix c.f. Sec.~\ref{sec_fisher}~\Luca{repetition}. 
%\eb{Justify the difference?}\Haowen{Added a few sentences} \Bei{Revised a bit.}
For each resolved event, we draw the recovered \ac{CBC} parameters %of the $i$th event 
from a multidimensional Gaussian distribution $\mathcal{N}(\widehat{\bm{\theta}}_\alpha,\Gamma_\alpha^{-1})$ with the mean located at the true values of the parameters $\widehat{\bm{\theta}}_\alpha$, and $\Gamma_\alpha$
given by Eq.~\eqref{eqs_detenet}.
%$\widehat{\bm{\Theta}}_i$ and the covariance matrix being the inverse Fisher matrix $\Gamma_i^{-1}$. 
We use \texttt{Bilby}~\cite{Ashton:2018jfp} again to generate the time series based on these recovered parameters. %from above processes.

Throughout this work, we define two types of masks, which we call \textit{ideal masks} and \textit{realistic masks}, respectively. 
The ideal masks are determined by \ac{CBC} injections with the true parameters, similarly to Ref.~\cite{Zhong:2022ylh}, while the realistic masks are instead determined by \ac{CBC} injections with the recovered parameters. In practice, realistic recovery due to \ac{PE} uncertainty implies that we can only have access to the realistic masks, but we consider the ideal masks as well for comparison purposes.  

To determine the masks, we consider a noise-free time series with resolvable \ac{CBC} injections only. We use the \texttt{Python} package $\texttt{pygwb}$~\cite{Renzini:2023qtj} to perform cross-correlation calculations and obtain $\hat{C}_{\mathrm{IJ}}(t_i; f_j)$ and $\hat{\sigma}_\mathrm{IJ}(t_i;f_j)$ for a given baseline IJ. From these two quantities, we construct \ac{SNR} maps as
\begin{equation}
    \mathrm{SNR}_\mathrm{IJ}(t_i;f_j)=\abs{\hat{C}_\mathrm{IJ}(t_i;f_j)/\hat{\sigma}_\mathrm{IJ}(t_i;f_j)}\,.
\end{equation}
We introduce an empirical \ac{SNR} threshold of $5\times 10^{-4}$ for identifying \ac{CBC} tracks, following the methodology detailed in Ref.~\cite{Zhong:2022ylh} for selecting an appropriate threshold. We note, however, that in our previous work, the threshold was set on $\hat{C}_\mathrm{IJ}(t_i;f_j)$, rather than SNR$_\mathrm{IJ}(t_i;f_j)$. The motivation for this change is that SNR$_\mathrm{IJ}(t_i;f_j)$ is more closely related to the \ac{SNR} for the stochastic search $\hat{C}_\mathrm{IJ}/\hat{\sigma}_\mathrm{IJ}$. 
%however, now we have switched to setting a threshold for SNR$_\mathrm{IJ}(t_i;f_j)$ instead. The motivation for this change is that SNR$_\mathrm{IJ}(t_i;f_j)$ is more closely related to SNR for the stochastic search $\hat{C}_\mathrm{IJ}/\hat{\sigma}_\mathrm{IJ}$. 
%~\Luca{My understanding is that the threshold for that work was chosen on $\sigma_\mathrm{IJ}$, not on SNR. I think we need a couple of sentences to explain the change and argue why.} \Bei{Why $10^{-3}$? I think the previous work chose the strain threshold $A_0$ by requiring $\hat{C}_{\rm IJ}/\hat{\sigma}_{\rm IJ} < 1$, right?}\Haowen{The way to define masks is not unique. We can set a threshold on SNR$(t_i,f_j)$ like here or on clean $\hat{C}_\mathrm{IJ}(t_i,f_j)$ like before. Also we did require $\hat{C}_\mathrm{IJ}/\hat{\sigma}_\mathrm{IJ}<1$ before, but this is the combined result instead of pixel value. In that case we set a threshold on $\hat{C}_{\mathrm{IJ}}(t;f)$ to calculate masks. Then we need to apply masks to all the spectrograms to make sure the final resulting $\hat{C}/\hat{\sigma}<1$. The reason to change from setting a threshold on $\hat{C}\mathrm{IJ}(t_i, f_j)$ to on SNR$(t_i, f_j)$ is because SNR of each pixel is more relevant to the SNR=$\hat{C}_\mathrm{IJ}/\hat{\sigma}_\mathrm{IJ}$.} 
%\Bei{Thanks for the explanation. But your previous paper wrote that the threshold was set on $A_0$, but maybe it's equivalent to the threshold on the $\hat{C}_\mathrm{IJ}(t_i;f_j)$ that you said here? And why choosing $\rm SNR=10^{-3}$ as the threshold?}
While the threshold can, in principle, vary across different baselines and can change over time and frequency, we simplify our approach by applying a uniform threshold across all baselines, time bins, and frequency bins.
We emphasize that, even if the time series used for cross correlation in this step are all noise free, we can still estimate $\hat{\sigma}_\mathrm{IJ}(t_i;f_j)$ according to Eq.~\eqref{eq:stoch} using the true PSD curves.

%\Luca{I think we should either move the discussion in the two paragraphs below to the appendix, or move the tables here.}

We summarize the cross-correlation results of noise-free (CBC only) data in Appendix ~\ref{app_tab} (see Tables~\ref{table:best_masks} and~\ref{table:biasedmasks}). In both cases, the masks are able to remove the resolvable \ac{CBC} foreground to a level below the detector noise, i.e., with SNR$\lesssim1$.

\begin{table*}[!htbp]
\centering
{\footnotesize
\begin{ruledtabular}
\begin{tabular}{
  l
  S[table-format=1.1e-2]
  S[table-format=1.1e-2]
  S[table-format=1.1e-2]
  S[table-format=1.1e-2]
  S[table-format=1.1e-2]
  S[table-format=1.1e-2]
  S[table-format=1.1e-2]
  S[table-format=1.1e-2]
  S[table-format=1.1e-2]
}
\toprule
\textbf{Baseline} & \multicolumn{3}{c}{\textbf{Before notching}} & \multicolumn{3}{c}{\textbf{Ideal masks applied}} & \multicolumn{3}{c}{\textbf{Realistic masks applied}} \\
\cmidrule(lr){2-4} \cmidrule(lr){5-7} \cmidrule(lr){8-10}
& {$\hat{C}_{\rm IJ}$} & {$\hat{\sigma}_{\rm IJ}$} & {SNR} & {$\hat{C}_{\rm IJ}$} & {$\hat{\sigma}_{\rm IJ}$} & {SNR} & {$\hat{C}_{\rm IJ}$} & {$\hat{\sigma}_{\rm IJ}$} & {SNR} \\
\midrule
H-L & 3.2e-10 & 1.4e-13 & 2.3e3 & 1.5e-11 & 1.6e-13 & 9.8e1 & 1.6e-11 & 1.5e-13 & 1.0e2 \\
H-ET1 & 3.1e-10 & 2.2e-12 & 1.4e2 & 3.2e-11 & 2.4e-12 & 1.3e1 & 3.4e-11 & 2.4e-12 & 1.4e1 \\
H-ET2 & 2.2e-10 & 1.2e-12 & 1.8e2 & 9.0e-12 & 1.8e-12 & 5.0 & 1.6e-11 & 1.8e-12 & 8.9 \\
H-ET3 & 2.7e-10 & 1.3e-12 & 2.1e2 & 1.5e-11 & 1.7e-12 & 8.8 & 1.9e-11 & 1.7e-12 & 1.1e1 \\
L-ET1 & 2.3e-10 & 1.5e-12 & 1.5e2 & 1.3e-11 & 2.3e-12 & 5.7 & 2.0e-11 & 2.3e-12 & 8.9 \\
L-ET2 & 2.5e-10 & 1.0e-12 & 2.5e2 & 1.1e-11 & 1.3e-12 & 7.9 & 1.5e-11 & 1.3e-12 & 1.2e1 \\
L-ET3 & 3.8e-10 & 2.2e-12 & 1.7e2 & 1.8e-11 & 2.4e-12 & 7.5 & 2.0e-11 & 2.4e-12 & 8.2 \\
ET1-ET2 & 3.1e-10 & 1.0e-12 & 3.0e2 & 2.5e-11 & 1.7e-12 & 1.5e1 & 4.1e-11 & 1.7e-12 & 2.5e1 \\
ET1-ET3 & 3.2e-10 & 1.1e-12 & 3.0e2 & 2.9e-11 & 1.7e-12 & 1.7e1 & 4.5e-11 & 1.7e-12 & 2.7e1 \\
ET2-ET3 & 3.2e-10 & 1.1e-12 & 3.0e2 & 2.8e-11 & 1.8e-12 & 1.6e1 & 4.4e-11 & 1.7e-12 & 2.5e1 \\
\midrule
Combined &3.1e-10&1.3e-13&2.4e3&1.5e-11&1.5e-13&1.0e2&1.7e-11&1.5e-13&1.1e2\\
\bottomrule
\end{tabular}
\end{ruledtabular}
}
\caption{Notching results for the Noise+CBC case. The power law index $\alpha$ is set to $\alpha=0$. The \ac{CBC} injections include all events, including the unresolvable ones.
Columns 2, 3 and 4 list $\hat{C}_\mathrm{IJ}$, $\hat{\sigma}_\mathrm{IJ}$ and SNR$=\hat{C}_\mathrm{IJ}/\hat{\sigma}_\mathrm{IJ}$ before notching. Columns 5, 6, and 7 are the results after applying ideal masks, which are determined by the true \ac{CBC} parameters. Columns 8, 9, and 10 list the results after applying realistic masks, calculated using the resampled \ac{CBC} parameters.}
\label{table:NC}
\end{table*}

\begin{table*}[!htbp]
\centering
{\footnotesize
\begin{ruledtabular}
\begin{tabular}{
  l
  S[table-format=1.1e-2]
  S[table-format=1.1e-2]
  S[table-format=1.1e-2]
  S[table-format=1.1e-2]
  S[table-format=1.1e-2]
  S[table-format=1.1e-2]
  S[table-format=1.1e-2]
  S[table-format=1.1e-2]
  S[table-format=1.1e-2]
}
\toprule
\textbf{Baseline} & \multicolumn{3}{c}{\textbf{Before notching}} & \multicolumn{3}{c}{\textbf{Ideal masks applied}} & \multicolumn{3}{c}{\textbf{Realistic masks applied}} \\
\cmidrule(lr){2-4} \cmidrule(lr){5-7} \cmidrule(lr){8-10}
& {$\hat{C}_{\rm IJ}$} & {$\hat{\sigma}_{\rm IJ}$} & {SNR} & {$\hat{C}_{\rm IJ}$} & {$\hat{\sigma}_{\rm IJ}$} & {SNR} & {$\hat{C}_{\rm IJ}$} & {$\hat{\sigma}_{\rm IJ}$} & {SNR} \\
\midrule
H-L & 3.6e-10 & 1.4e-13 & 2.6e3 & 5.5e-11 & 1.6e-13 & 3.5e2 & 5.6e-11 & 1.5e-13 & 3.6e2 \\
H-ET1 & 3.5e-10 & 2.2e-12 & 1.6e2 & 7.1e-11 & 2.4e-12 & 2.9e1 & 7.4e-11 & 2.4e-12 & 3.1e1 \\
H-ET2 & 2.6e-10 & 1.2e-12 & 2.1e2 & 4.9e-12 & 1.8e-12 & 2.7e1 & 5.6e-11 & 1.8e-12 & 3.1e1 \\
H-ET3 & 3.1e-10 & 1.3e-12 & 2.4e2 & 5.5e-11 & 1.7e-12 & 3.2e1 & 5.8e-11 & 1.7e-12 & 3.5e1 \\
L-ET1 & 2.7e-10 & 1.5e-12 & 1.8e2 & 5.3e-11 & 2.3e-12 & 2.3e1 & 6.0e-11 & 2.3e-12 & 2.6e1 \\
L-ET2 & 2.9e-10 & 1.0e-12 & 2.9e2 & 5.1e-11 & 1.3e-12 & 3.8e1 & 5.5e-11 & 1.3e-12 & 4.2e1 \\
L-ET3 & 4.2e-10 & 2.2e-12 & 1.9e2 & 5.8e-11 & 2.4e-12 & 2.4e1 & 6.0e-11 & 2.4e-12 & 2.5e1 \\
ET1-ET2 & 3.4e-10 & 1.0e-12 & 3.3e2 & 6.5e-11 & 1.7e-12 & 3.8e1 & 8.1e-11 & 1.7e-12 & 4.9e1 \\
ET1-ET3 & 3.6e-10 & 1.1e-12 & 3.4e2 & 6.9e-11 & 1.7e-12 & 3.9e1 & 8.5e-11 & 1.7e-12 & 5.0e1 \\
ET2-ET3 & 3.6e-10 & 1.1e-12 & 3.4e2 & 6.7e-11 & 1.8e-12 & 3.8e1 & 8.3e-11 & 1.7e-12 & 4.8e1 \\
\midrule
Combined &3.5e-10&1.3e-13&2.8e3&5.5e-11&1.5e-13&3.7e2&5.7e-11&1.5e-13&3.8e2\\
\bottomrule
\end{tabular}
\end{ruledtabular}
}
\caption{Same as Table~\ref{table:NC}, but for the Noise+CBC+SGWB case.}
\label{table:NCS}
\end{table*}

% \begin{table*}[!htbp]
% \centering
% \begin{ruledtabular}
% \begin{tabular}{
%   l
%   c
%   S[table-format=1.1e-2]
%   S[table-format=1.1e-2]
%   S[table-format=1.1e-2]
% }
% \toprule
% &{\textbf{Cases}} & {\textbf{Before notching}} & {\textbf{Ideal masks applied}} & {\textbf{Realistic masks applied}} \\
% \midrule
% \multirow{2}{*}{$\OGW^{5-200 \mathrm{Hz}}$}&Noise+CBC & 2.2e-09 & 2.0e-10 & 2.8e-10 \\
% &Noise+CBC+SGWB & 2.4e-09 & 3.5e-10 & 4.2e-10 \\
% \bottomrule
% \end{tabular}
% \end{ruledtabular}
% \caption{$\Omega_\mathrm{GW}^{5-200 \mathrm{Hz}}$ accumulated across all frequency bins and baselines for the Noise+CBC and Noise+CBC+SGWB cases. We show the results before notching, after applying ideal masks, and after applying realistic masks. In all cases, the energy density of the astrophysical foreground decreases by about one order of magnitude when we apply the masks.}
% \label{table:OMega_GW_Cases}
% \end{table*}

\section{Notching Results}\label{sec_notching}

After computing the masks with noise-free data, we now apply them to noise injections to test their performance.

First, we consider the case where we simulate detector noise and all \ac{CBC} events in our catalog, including the unresolvable ones. For simplicity, we call this scenario the ``Noise+CBC case''. Then we consider a scenario in which we also simulate a flat ($\alpha=0$) \ac{SGWB} with amplitude $\Omega_\alpha=4\times 10^{-11}$ in addition to the Noise+CBC data.  %and repeat the notching procedure. %to get Table~\ref{table:NCS}. 
We call this the ``Noise+CBC+SGWB case''. We note that the real cosmological \ac{SGWB} is not necessarily flat. Here we choose this specific \ac{SGWB} only for demonstration purposes. The results of the notching procedure for two scenarios are shown in Table~\ref{table:NC} and Table~\ref{table:NCS}, and discussed in Sec.~\ref{sec_NC&NCS} below.

After generating the time series we run \texttt{pygwb} to get $\hat{C}_\mathrm{IJ}(t_i;f_j)$ and $\hat{\sigma}_\mathrm{IJ}(t_i; f_j)$, assuming $\alpha=0$ in both cases. We then apply ideal masks and realistic masks upon the time-frequency maps to remove the astrophysical foreground produced by the resolved \ac{CBC} events. The injection of the \ac{SGWB} is also done with \texttt{pygwb}. Both tables show the results before notching, and after applying ideal masks and realistic masks for all the different baselines. The HL baseline has the highest sensitivity for searches of \acp{SGWB} with power-law index $\alpha=0$. This is why the results found by combining all ten baselines are almost identical to the results given by the single HL baseline.

\subsection{Results of the notching procedure}
%\subsection{Noise+CBC and Noise+CBC+SGWB cases}
\label{sec_NC&NCS}
In the Noise+CBC case (Table~\ref{table:NC}), there are only two remaining components in the data after notching, in addition to detector noise: (i) The \ac{CBC} foreground from individually unresolved events $\hat{C}_\mathrm{unresolvable}$, which cannot be further removed; this is effectively the only remaining component present in the ideal mask case~\cite{Zhong:2022ylh}; (ii) The residual foreground due to realistic notching $\hat{C}_\mathrm{residual}$; this is an additional component which is present when \ac{PE} uncertainties are taken into account by using realistic masks, due to the residuals from realistic subtraction of resolved signals.

From the last row of Table~\ref{table:NC} we see that the resulting $\hat{C}$ is reduced from $3.1\times 10^{-10}$ to $1.5\times 10^{-11}$ after applying the ideal masks, and to $1.7(1.68)\times 10^{-11}$ after applying the realistic masks. In other words, our notching procedure can still suppress the foreground from all \ac
{CBC} events to $\sim5.4\%$ of its original level, even when only resolvable \ac{CBC} events are removed and the residuals due to realistic subtraction are taken into account. This is one of the main results of this paper.

From the ideal-mask case, we obtain that $\hat{C}_\mathrm{unresolvable}\approx1.5(1.54)\times 10^{-11}$, and thus the additional contribution with realistic masks is $\hat{C}_\mathrm{residual}\approx1.4(1.44)\times 10^{-12}$. The ratio between the residual foreground and the foreground from unresolvable events is $\hat{C}_\mathrm{residual}/\hat{C}_\mathrm{unresolvable}\approx 9.3\%$. This implies that, despite the additional contribution from realistic removal, the floor for \ac{SGWB} searches is largely determined by the unresolvable \ac{CBC} foreground, and not by the residual due to notching. This is consistent with previous conclusions in the idealized case of perfect recovery of the resolved signals~\cite{Zhong:2022ylh}.
\begin{figure*}[t]
\centering
\includegraphics[width=1.0\textwidth]{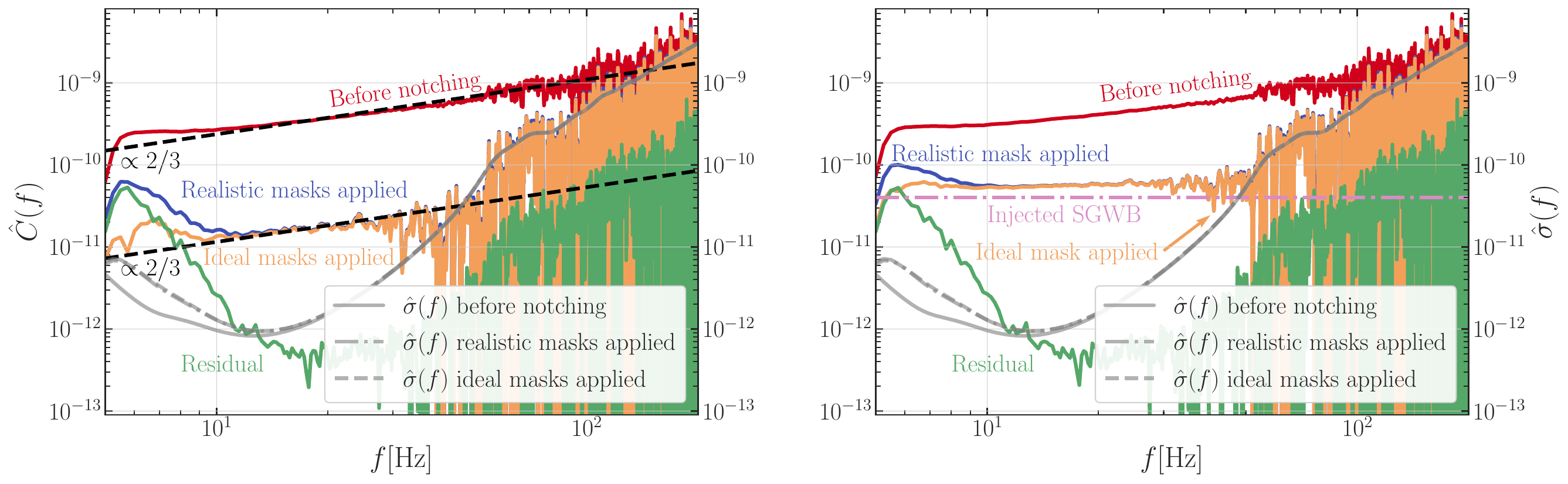}
\caption{Left panel: combined $\hat{C}(f)$ and $\hat{\sigma}(f)$ for the Noise+CBC case. Right panel: same, but for the Noise+CBC+SGWB case (the injected \ac{SGWB} is shown by the dot-dashed pink curve). The red, orange, and blue curves correspond to the results before notching, after applying ideal masks, and after applying realistic masks, respectively. The curve labeled as “residual” (in green) is the difference between the blue and orange curves. The combined sensitivity $\hat{\sigma}(f)$ is plotted in gray: the solid, dashed, and dot-dashed gray curves correspond to the sensitivity before notching, after applying ideal masks, and after applying realistic masks, respectively.  \\}
\label{fig:NC&NCS}
\end{figure*}
%In this scenario, we note that there are only two remaining components in the data after notching, on top of detector noise: (i) the \ac{CBC} foreground from individually unresolvable events, which cannot be further removed and (ii) the residual foreground due to realistic notching. 

%According to Table~\ref{table:best_masks}, we can tell that after applying the ideal masks, the residual foreground due to notching is negligible, so the remaining $\hat{C}=1.5\times 10^{-11}\approx\hat{C}_\mathrm{unresolvable}$ is almost totally due to unresolvable \ac{CBC} foreground. While for the case where realistic masks are applied, we cannot remove all the resolvable foreground perfectly. Hence, the resulting $\hat{C}=1.7\times 10^{-11}$ is composed by $\hat{C}_\mathrm{unresolvable}\approx1.5\times 10^{-11}$ and $\hat{C}_\mathrm{residual}\approx 1.4\times 10^{-12}$. The ratio between the residual foreground $\hat{C}_\mathrm{residual}$ and unresolvable foreground reads $\hat{C}_\mathrm{residual}/\hat{C}_\mathrm{unresolvable}\approx 9.3\%$. This finding again implies that the floor of the \ac{SGWB} search is actually determined by the unresolvable \ac{CBC} foreground instead of residual from notching which is consistent with our previous conclusion~\cite{Zhong:2022ylh}.

The last row of Table~\ref{table:NC} shows that the combined $\hat{\sigma}$ is $1.3\times 10^{-13}$ before notching, and it increases to $1.5\times10^{-13}$ when we apply either ideal or realistic masks. It is important to note that $\hat{\sigma}$ increases by $16\%$ after notching, which means that the sensitivity for stochastic searches will be moderately impacted by the foreground cleaning procedure. This is also consistent with the conclusions drawn in Ref.~\cite{Zhong:2022ylh}. 

%The numbers we report in Table~\ref{table:NC} and Table~\ref{table:NCS} are dependent on the choice of $\alpha$. To show an $\alpha$-independent estimator, we accumulate $\OGW(f)$ across all frequency bins (5-200 Hz) and all baselines to get $\OGW^{5-200 \mathrm{Hz}}$~\Luca{How is $\OGW(f)$ inferred from data here? Is it from inverting Eq. 5? Maybe we should specify, the only definition given so far is the power law of Eq. 2.}:
%\begin{equation}
%    \OGW^{5-200~\mathrm{Hz}}:=\int_{5~\mathrm{Hz}}^{200~\mathrm{Hz}} \OGW(f)~\D\ln f.
%    \label{eq:GW_accumulate}
%\end{equation}
%\Luca{Move this definition above?}
%We report $\OGW^{5-200 \mathrm{Hz}}$ in Table~\ref{table:OMega_GW_Cases}. 
%The second to forth columns correspond to $\OGW^{5-200 \mathrm{Hz}}$ under three scenarios, where no mask is applied, ideal masks are applied and realistic masks are applied. 

% In the first row of Table~\ref{table:OMega_GW_Cases} we show the $\alpha$-independent estimator of Eq.~\eqref{eq:GW_accumulate} for the Noise+CBC case. We find that the original foreground $\OGW^{5-200 \mathrm{Hz}}=2.2\times 10^{-9}$ before notching is successfully suppressed to $2.0\times 10^{-10}$ after applying ideal masks, and to $2.8\times10^{-10}$ after applying realistic masks. The remaining foreground after notching is $\sim 40\%$ larger when \ac{PE} uncertainties are taken into account, but it is still suppressed by $\sim 90\%$ compared to the original value before notching.

%we still can decrease $\OGW^{5-200 \mathrm{Hz}}$ by a order of magnitude to $2.8\times10^{-10}$.

Table~\ref{table:NC} lists the values of $\hat{C}_\mathrm{IJ}$ and $\hat{\sigma}_\mathrm{IJ}$ combined %by $\hat{C}_\mathrm{IJ}(f)$ and $\hat{\sigma}_\mathrm{IJ}(f)$ 
across all frequency bins. In Fig.~\ref{fig:NC&NCS} we also show $\hat{C}(f)$ and $\hat{\sigma}(f)$ as functions of frequency, when we combine all ten baselines. The red, blue\footnote{The turning point of the red and blue curves around $\sim6~\rm{Hz}$ turns out to be a numerical artifact. When injecting \ac{CBC} signals into the time series, we need to set a minimum frequency, which we choose to be $f_{\mathrm{min}}=5~\rm{Hz}$.  Also we note that when applying cross-correlation using \texttt{pygwb}, a cutoff frequency needs to be specified. We also set $f_\mathrm{cut}=5$ Hz in the analysis. Had we set lower minimum and cutoff frequency the turning point in $\hat{C}(f)$ around 6 Hz would not be present. However, setting a lower minimum frequency would result in much longer computational times. We note that the most sensitive frequency range for \ac{SGWB} searches is around $\mathcal{O}$ (10) Hz, so the impact of the chosen frequency cutoff on the resulting $\hat{C}$ is minimal.} and orange curves show $\hat{C}(f)$ before applying masks, after applying realistic masks, and after applying ideal masks, respectively. We estimate the residual from realistic subtraction $\hat{C}_{\rm residual}$ (shown in green) by taking the difference between the blue and orange curves. The combined sensitivity $\hat{\sigma}(f)$ in the three cases (before notching, ideal masks, and realistic masks) are overplotted in gray with different line styles (solid, dashed, and dot-dashed, respectively). 

The curves for ideal and realistic masks %Orange curve and blue curve 
are in remarkable agreement within the frequency range $f\sim 10-50$ Hz, which coincides with the most sensitive band of the stochastic search for $\alpha=0$. This illustrates the capability of realistic masks to remove resolvable \ac{CBC} signals with high accuracy. The residual foreground is relatively loud in the low-frequency range $f\sim 5-10$ Hz, because \ac{CBC} signals overlap significantly in this range~\cite{Johnson:2024foj}. Comparing to Fig.~2 of Ref.~\cite{Zhong:2022ylh}, we notice that nearly all the pixels at those low frequencies are notched out. Therefore, a slight bias in the masks coming from \ac{PE} uncertainties would result in a relatively loud residual.

We highlight that in the frequency range $f\gtrsim 10$ Hz, the green curve (residual foreground) is well below the gray curves (detector noise). This implies that the residual foreground due to realistic notching has minimal impact on the search for the cosmological \ac{SGWB}. Furthermore, we emphasize that neither the gray curves nor the green curve limit the search for the cosmological \ac{SGWB}. \textit{The real limit on the detectability of the cosmological \ac{SGWB} is set by the blue curve}, which represents the superposition of the unresolvable foreground and of the residual foreground. As we discussed, the unresolvable foreground is about ten times stronger than the residual foreground. This confirms that the unresolvable \ac{CBC} foreground is the main obstacle in the search for the cosmological \ac{SGWB}.
%(Here we naively compare the relative strength of the cosmological background and the remaining foreground after notching: if the cosmological background is stronger than the remaining foreground, then we call it ``detectable.'')
In Table~\ref{table:NCS} and in the right panel of Fig.~\ref{fig:NC&NCS}, we show the same results as Table~\ref{table:NC} and the left panel of Fig.~\ref{fig:NC&NCS}, but with an additional flat ($\alpha=0$) stochastic background with $\Omega_\mathrm{ref}=4\times 10^{-11}$ injected into the data.  As mentioned above, the astrophysical foreground is suppressed by an order of magnitude, but the strength of the remaining astrophysical foreground is comparable to the injected cosmological \ac{SGWB}. Therefore, even though the \ac{SNR} is as large as $3.8\times 10^2$, we still cannot claim a convincing detection of the cosmological \ac{SGWB}. Better methods will be required to separate the remaining astrophysical foreground from the cosmological background of interest. This is an interesting topic for future work.
\begin{figure}[t]
    \centering
    \includegraphics[width=0.49\textwidth]{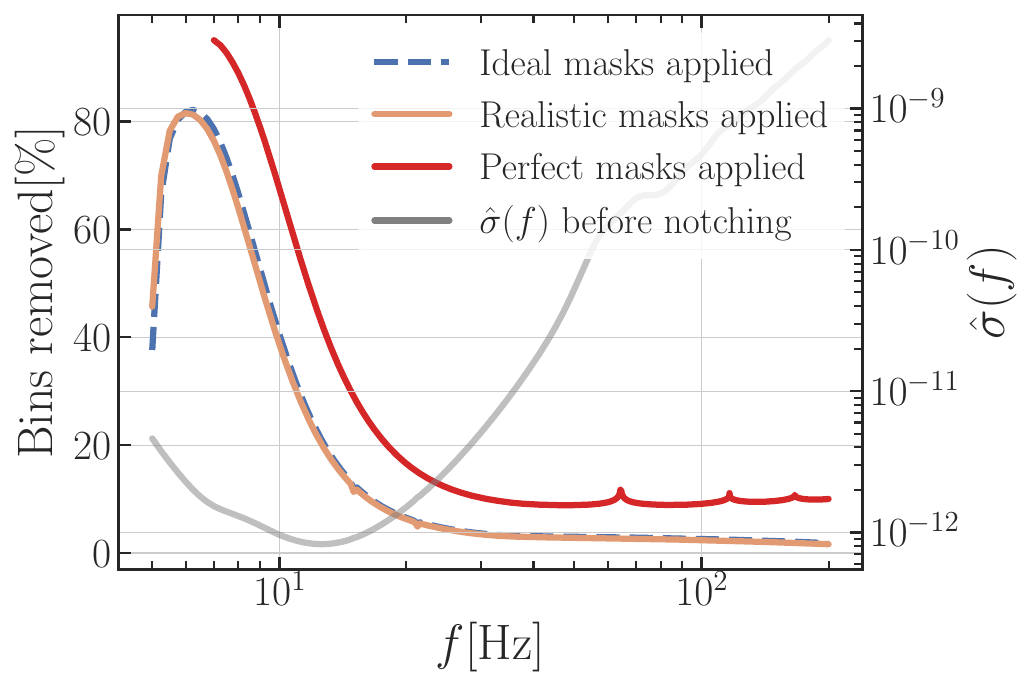}
    \caption{Percentage of bins removed as a function of frequency. The red curve is the orange curve in Fig.~2 of Ref.~\cite{Zhong:2022ylh}, which characterizes the proportion of frequencies bins removed by the \textit{perfect masks}, which remove all \ac{CBC} events including unresolvable ones. The orange curve and blue dashed curve correspond to the results of the realistic masks and ideal masks, respectively. The two curves coincide quite well. For reference, the combined sensitivity $\hat{\sigma}(f)$ before notching is plotted in gray.}
    \label{fig:Percentage}
\end{figure}

\subsection{Percentage of bins removed and missing rate}

We now assess in more detail the performance of the realistic masks relative to the ideal masks.
%\Luca{For the \ac{CBC}+Noise case, I assume?}\Haowen{This discussion is not specific for CBC+Noise case or CBC+Noise+SGWB case. Since we are just comparing two sets of masks.}

In Fig.~\ref{fig:Percentage} we show the percentage of bins removed as a function of frequency. The red curve is exactly the same as the orange curve in Fig.~2 of Ref.~\cite{Zhong:2022ylh}, %In that case, 
where we assumed we could detect all \ac{CBC} events (including the unresolvable ones) and recover \ac{CBC} parameters perfectly. 
The yellow and blue curves correspond to the results after applying realistic masks and ideal masks, respectively. The gray curve shows the sensitivity $\hat{\sigma}(f)$ before notching. The difference in starting frequencies between the red curve and the other two stems from the different choices of $f_\mathrm{min}$ in the two analyses. %We highlight a methodological shift in computing masks between this study and our prior work~\cite{Zhong:2022ylh}. Previously, the threshold for notching was applied directly to clean $\hat{C}_\mathrm{IJ}(t_i;f_j)$ maps. In contrast, the current approach updates the masking process by employing a threshold on SNR maps instead.~\Luca{Can we elaborate a bit on why the change in method? Maybe move this to SEC. 5 when we introduce the masks?}. 
As remarked in Ref.~\cite{Zhong:2022ylh}, the three peaks in the red curve can be attributed to the zeros in the overlap reduction function $\gamma_\mathrm{IJ}$ that are specific to the HL baseline. Transitioning to an \ac{SNR}-based threshold and incorporating a wider array of baselines transforms these peaks into dips, with a corresponding shift in their locations.

The yellow and blue curves nearly overlap with each other. This confirms once again the similar performance of realistic masks and ideal masks discussed above. The qualitative behavior of the three curves as a function of frequency is similar, but the orange and blue curves are both well below the red curve in the whole frequency band. This is because in this work, we distinguish between resolvable and unresolvable signals, and the pixels containing the latter component
% have no way to remove those unresolvable \ac{CBC} events and those pixels containing unresolvable \ac{CBC} signals will finally 
survive even after the notching procedure.

%\begin{figure}[!htbp]
\begin{figure}[t]
    \centering
    \includegraphics[width=0.49\textwidth]{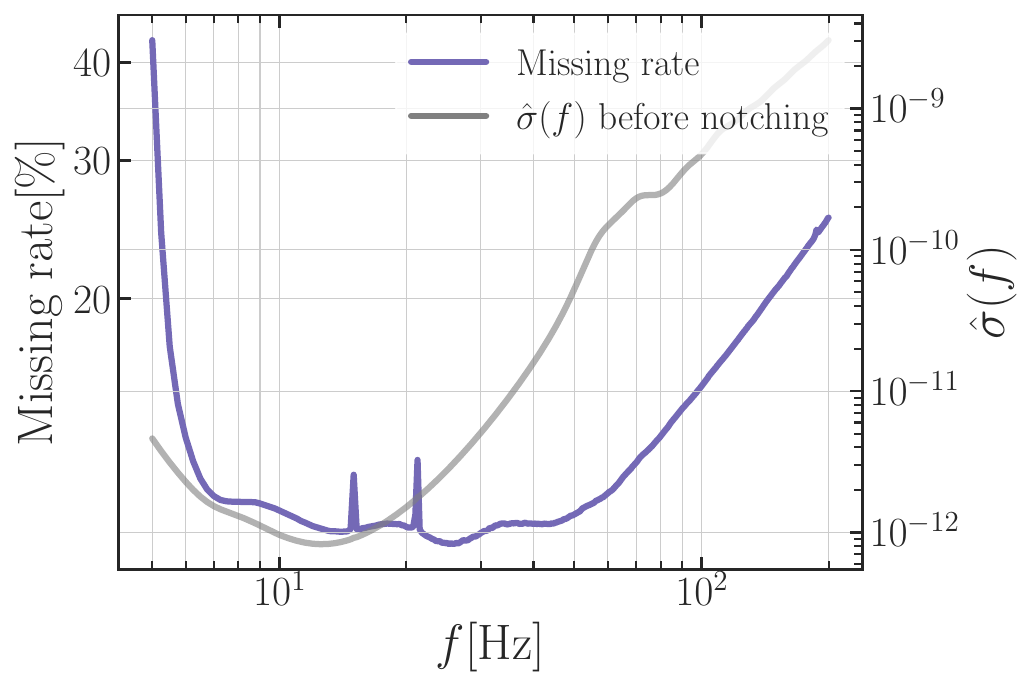}
    \caption{Missing rate as a function of frequency. The missing rate is defined as the average percentage of pixels that are removed by the ideal masks yet kept by the realistic masks, relative to the total number of pixels removed by the ideal masks. For reference, the combined sensitivity $\hat{\sigma}(f)$ before notching is plotted in gray.}
    \label{fig:Missing_rate}
\end{figure}

For a specific frequency, we can define a ``missing rate'' as the average
%\footnote{Our injections are saved by the format of \ac{GW} frames. We calculate the ideal mask and realistic mask for each frame. The missing rate is calculated for the mask corresponding to each frame and then averaged over all frames. \eb{I don't get this.}}
percentage %\eb{average over what?} \Haowen{added a footnote}%\Bei{proportion $\rightarrow$ number?}
of pixels that are removed by the ideal masks yet kept by the realistic masks, $\Delta N(f)$, normalized to the total number of pixels removed by the ideal masks, $\hat{N}(f)$:

\begin{equation}
    \mathrm{Missing~Rate}(f)=\Bigg\langle\frac{\Delta N(f)}{\hat{N}(f)}\Bigg\rangle_\mathrm{masks} .
\end{equation}

In Fig.~\ref{fig:Missing_rate} we plot the missing rate as a function of frequency. For reference, we also plot the combined sensitivity $\hat{\sigma}(f)$ before notching. The missing rate is relatively high at low and high frequencies, while it reaches a minimum value in the frequency range $f\sim 10-100~\rm{Hz}$, where our search is most sensitive, and therefore the shape of \ac{CBC} tracks in the time-frequency domain is best determined. Since we miss the fewest pixels in our most sensitive band, the resulting $\hat{C}$ after applying realistic masks should be rather close to $\hat{C}$ after applying ideal masks in this frequency band. These findings are consistent with those in Table~\ref{table:NC}, Table~\ref{table:NCS} and Fig.~\ref{fig:NC&NCS}.

\subsection{Physical interpretation}

From the above discussion, it is clear that the notching procedure is still effective at suppressing the foreground from resolved signals even when \ac{PE} uncertainties are taken into account. These findings are significantly more optimistic than previous work by some of us~\cite{Zhou:2022nmt, Zhou:2022otw}, which estimated the residual foreground from realistic subtraction in the frequency domain to be comparable to the original foreground.

To understand the reason for this improvement, it is useful to think about which binary parameters are mainly responsible for realistic subtraction in the two different approaches. In the analysis of Ref.~\cite{Zhou:2022nmt, Zhou:2022otw}, the residuals are dominated by the errors in luminosity distance and coalescence phase (which, in turn, is strongly correlated with other extrinsic parameters, such as the polarization angle). These parameters are poorly constrained for a significant fraction of events even with a network of \ac{XG} detectors.
%To address the second question, we turn to the frequency evolution of a binary system during its inspiral phase. 
By contrast, the performance of a mask in the notching procedure is largely determined by how well we can recover the location of the time-frequency track for each event. 
%According to Newtonian approximation, we have~\cite{foft}:
At leading (quadrupolar) order, the time-frequency evolution for a \ac{CBC} in the inspiral phase is set by~\cite{Peters:1963ux}
\begin{equation}
    f(t,t_c;\mathcal{M}_c^{z})=\frac{1}{\pi}\Big(\frac{5}{256(t_c-t)}\Big)^{\frac{3}{8}}\Big(\frac{G\mathcal{M}_c^z}{c^3}\Big)^{-\frac{5}{8}}.
\end{equation}
Given that most events in \ac{XG} detectors are inspiral-dominated, this approximation works well for the vast majority of our catalogs.
%where $t_c$ is the coalesence time, and $\mathcal{M}_c^z=(1+z)\mathcal{M}_c$ is the detector frame chirp mass. 
%In the frequency range of our interest $f\sim 5-200$ Hz, the Newtonian approximation is sufficient. 
Hence, the location of a \ac{CBC} track in the time-frequency plane is mostly determined by the detector-frame chirp mass $\mathcal{M}_c^z$. Given an accurate recovery of $\mathcal{M}_c^z$, the masks computed according to biased injections can still be effective at removing astrophysical foregrounds. 

\begin{figure}[t]
    \centering
    \includegraphics[width=0.49\textwidth]{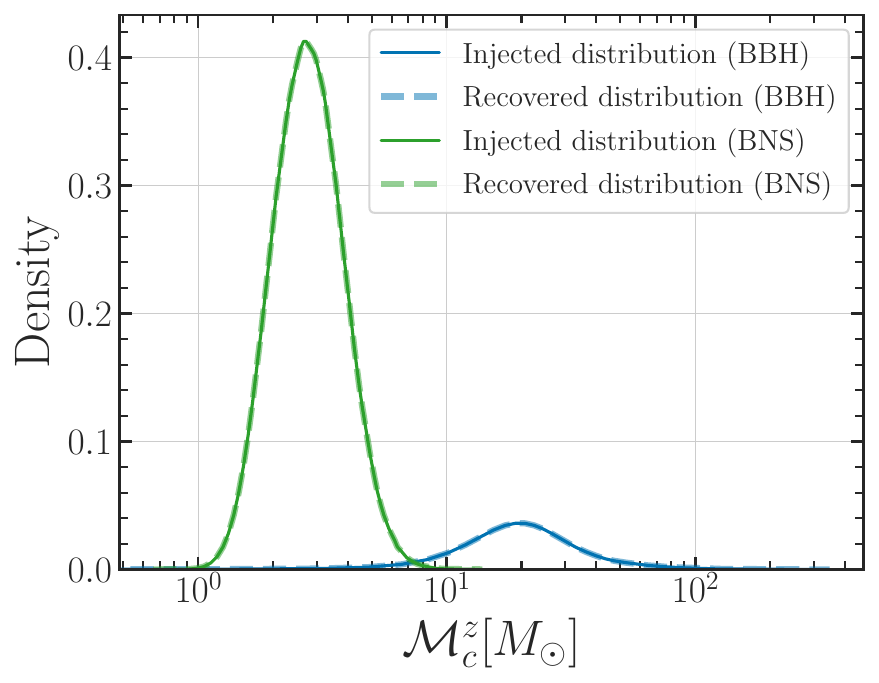}
    \caption{Injected and recovered distribution of detector-frame chirp masses, $\mathcal{M}_c^z$, for \ac{BBH} and \ac{BNS} systems. The recovered distribution coincides with the injected distribution quite well.}
    \label{fig:M_cz}
\end{figure}

To confirm this interpretation, in Fig.~\ref{fig:M_cz}, we compare the injected and recovered distributions of detector-frame chirp masses for our \ac{BBH} and \ac{BNS} catalogs, respectively. %The blue histograms show the injected $\mathcal{M}_c^z$ distribution, while the orange histograms show the recovered $\mathcal{M}_c^z$ distribution. 
The two histograms show remarkable agreement for both \acp{BBH} and \acp{BNS}, because the chirp mass is constrained extremely well with \ac{XG} observatories~\cite{Borhanian:2022czq, Iacovelli:2022bbs, Pieroni:2022bbh}. This precision guarantees that even if the amplitude of individual events is recovered poorly, we can still accurately determine the \ac{CBC} chirps in the time-frequency domain, such that the notching procedure is effective.

%the stochastic background amplitude is biased due to poor recovery of luminosity distance for individual event, we are still able to determine the positions of \ac{CBC} chirps in the time-frequency domain with sufficient accuracy.

%~\Luca{Expand discussion, comparing to our previous paper. Discuss DL, phase, etc.}
%\begin{figure}[!htbp]

%%%%%%%%%%%%%%%%%%%%%%%%%%%%%%%%%%%
%%%%%%       SECTION        %%%%%%%
%%%%%%%%%%%%%%%%%%%%%%%%%%%%%%%%%%%
\section{Discussion and conclusions}\label{sec_dis}

In this work, we improve the notching method of Ref.~\cite{Zhong:2022ylh} by extending it to a more realistic scenario where not all \ac{CBC} signals are resolved and where \ac{PE} uncertainties on the resolved signals are taken into account. We generate \ac{BBH} and \ac{BNS} catalogs consistent with the observed \rm{LVK} populations, and for each resolved event, we estimate the errors on the inferred parameters with a Fisher matrix approach. We compute realistic masks on the recovered \ac{CBC} signals by resampling their parameters while taking into account the corresponding uncertainties. We apply these realistic masks to notch out the resolved signals in the time-frequency domain. We then compare the result of cross-correlation \ac{SGWB} searches after this notching procedure with the previous case, where ideal masks were computed by using the true parameters of the injected signals.

We find that the realistic masks perform remarkably well. Consistent with the results of Ref.~\cite{Zhou:2022nmt, Zhou:2022otw}, the amplitude of the foreground in the frequency domain from the resampled parameters can be significantly biased. The poor recovery of luminosity distance, along with the fact that close-by events dominate the foreground, leads to a \ac{CBC} foreground before notching that is significantly larger when \ac{PE} errors are taken into account (see Tables~\ref{table:best_masks} and~\ref{table:biasedmasks}). However, the tracks in the time-frequency plane are mainly determined by the detector-frame chirp mass $\mathcal{M}_c^z$, which is constrained to high precision for most events, thus the notching procedure is still very effective at suppressing the \ac{CBC} foreground.

Assuming a flat ($\alpha=0$) power-law search, we find that the residual \ac{CBC} foreground $\hat{C}$ due to realistic notching is $\sim 10\%$ of the foreground from unresolved signals that cannot be removed. With our assumptions on the \ac{BBH} and \ac{BNS} merger rate, we find that the total foreground is reduced to about $\sim 5.4\%$ of its original value. We note that the reported numbers are all based on our assumptions about the \ac{CBC} population. If a different population were assumed, the method we proposed in Ref.~\cite{Zhong:2022ylh} would still allow us to determine the optimal value of the threshold and compute masks for notching. Even though the strength of the residual foreground and unresolvable \ac{BNS} foreground can vary due to different assumptions on the \ac{BNS} rate, the relative strengths of these two contributions should depend only weakly on the local \ac{BNS} merger rate.
% Adopting a power-law independent estimator  $\OGW^{5-200~\mathrm{Hz}}$, we find that the residual foreground when realistic recovery is taken into account is $\sim 40\%$ larger than the idealized case of perfect recovery. However, the total foreground after notching is still suppressed by $\sim 90\%$ compared to the original value.

%due to our ability to recover the detector frame chirp mass $\mathcal{M}_c^z$ of \ac{CBC} systems accurately, the realistic masks still can perform well. The total foreground is suppressed to its original $5.4\%$ level after notching and the residual foreground due to realistic notching is limited to the $9.3\%$ level of unresolvable \ac{CBC} foreground.

%Fig.~\ref{fig:NC}, Fig.~\ref{fig:NCS}, Table~\ref{table:NC}, and Table~\ref{table:NCS} all imply that the the notching procedure proposed in Ref.~\cite{Zhong:2022ylh} is able to reach the limit set by unresolvable \ac{CBC} events. When uncertainties in the stage of \ac{PE} is considered, the direct estimation of resulting $\hat{C}^\mathrm{recovered}(f)$ might be strongly deviated from the true $\hat{C}^\mathrm{true}(f)$. However, due to our ability to recover the detector frame chirp mass $\mathcal{M}_c^z$ of \ac{CBC} systems accurately, the realistic masks still can perform well. The total foreground is suppressed to its original $5.4\%$ level after notching and the residual foreground due to realistic notching is limited to the $9.3\%$ level of unresolvable \ac{CBC} foreground. 

It is important to point out some caveats about the notching procedure used here, and some directions in which it could be further improved. 
We only resample \ac{CBC} parameters %from $\mathcal{N}(\hat{\bm{\Theta}},\Gamma^{-1})$ 
once for each resolvable event, but multiple draws could be made for low-SNR events to better remove their tracks in the time-frequency plane. However, this also means notching out more pixels, which reduces the detectability of a subdominant \ac{SGWB}. A more detailed study is needed to understand whether multiple draws would be effective. %\Bei{Multidraw has a downside, which is less survived pixels. I think this should be mentioned.}
We assume that the detector noise is perfectly known, while in practice estimating the detector \ac{PSD} in the presence of glitches, non-stationary noise, and the foreground from unresolved \ac{CBC} events can be challenging. 
We neglect the impact of waveform systematics, which is subdominant compared to current astrophysical uncertainties, but still relevant for correctly estimating the foreground~\cite{Zhou:2022nmt, Zhou:2022otw}. 
Finally, we neglect astrophysical uncertainties in the \ac{CBC} catalogs. In particular, we only consider one possible value of the \ac{BNS} merger rate, which is within the large uncertainty of current \ac{LVK} catalogs. If the number of unresolved events is larger, either due to higher merger rates overall or larger contributions at higher redshifts (e.g.~\cite{Boesky:2024msm, Boesky:2024wks}), the \ac{CBC} foreground after notching may be significantly higher.
However, this would not affect the main conclusion of this work -- i.e., that including uncertainties in the binary parameters does not significantly affect the notching method.
% \Bei{Yes, but I think this wouldn't affect the main conclusion of this paper, i.e., including PE barely affects the efficacy of the notching method? If so we may need to add a sentence to clarify.}
Finally, we neglect the contribution of \ac{NSBH} binaries, which can produce an unresolved foreground comparable to the one from \acp{BNS} in \ac{XG} detectors~\cite{Bellie:2023jlq}. All of these limitations should be addressed in future work.
%\Bei{This paragraph listed a lot of assumptions or things we didn't do in our calculation, and the wording makes our paper look not so good. Actually, I think most, if not all, of these points affect the major conclusion our paper, i.e., including PE does not affect the efficacy of the notching method. So I'd suggest tweaking the wording/tone of this paragraph a bit.}

\begin{figure*}[t]
    \centering
    \includegraphics[width=0.49\textwidth]{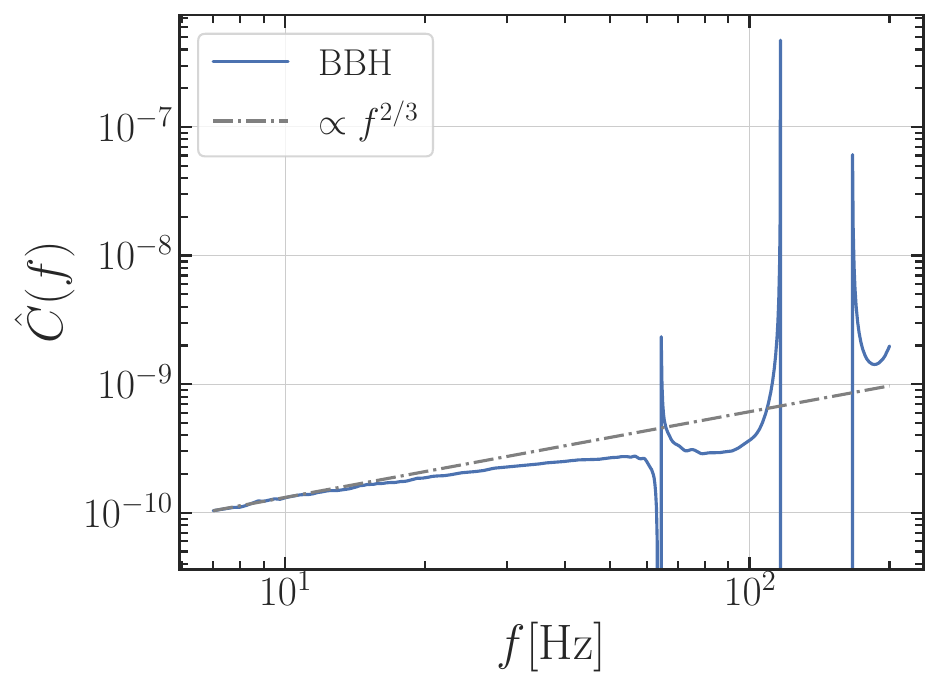}
    \includegraphics[width=0.49\textwidth]{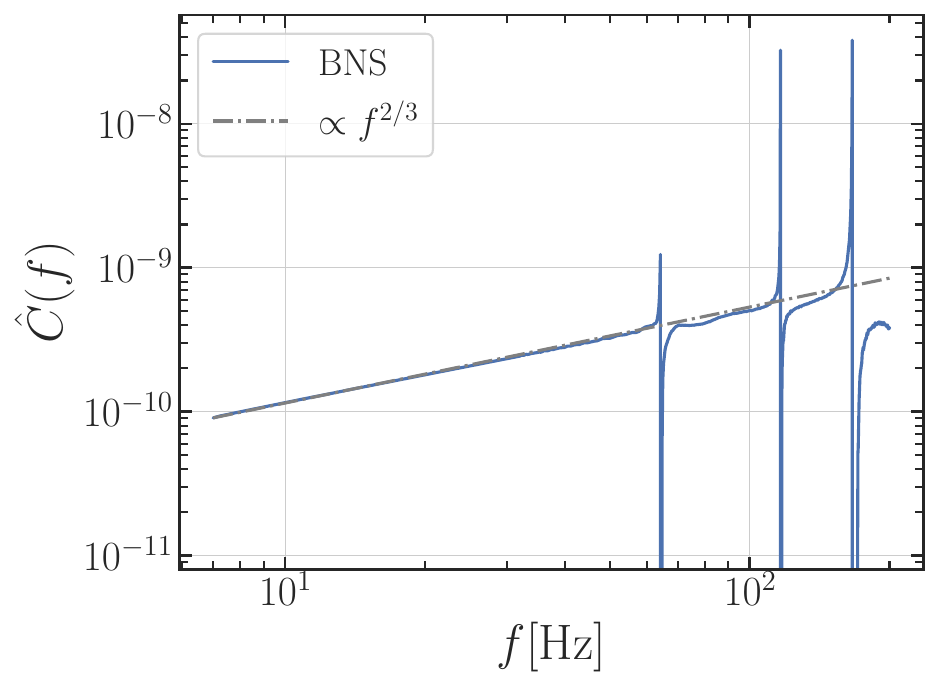}
    \caption{$\hat{C}(f)$ for \ac{BBH} and \ac{BNS} systems. The gray dotted-dashed lines are included for comparison and represent the $f^{2/3}$ power-law, showcasing the typical gravitational wave frequency evolution characteristic of compact binary inspirals. The spectral behavior of $\hat{C}(f)$ for \ac{BBH} systems exhibits a notable deviation from the $f^{2/3}$ power-law at frequencies greater than 20 Hz. In contrast, for \ac{BNS} systems, $\hat{C}(f)$ aligns closely with the $f^{2/3}$ power-law across the entire frequency band of interest. The peaks in the curves are due to zero points in the overlap reduction function (see. Eq.~\eqref{eq:stoch}).}
    \label{fig:simple_test}
\end{figure*}

Overall, our results are consistent with Ref.~\cite{Zhong:2022ylh}, which suggests that the limiting factor in the sensitivity of \ac{XG} detectors to stochastic searches will be the \ac{CBC} foreground from individually unresolved signals, which cannot be further notched out. It is important to develop techniques to separate different \ac{SGWB} components from the unresolved \ac{CBC} foreground after notching, which in turn requires accurate modeling of the \ac{CBC} foreground from unresolved events. 

Several techniques to address this issue can and should be explored in the future. For instance, template-based methods that simultaneously fit for both resolved and unresolved signals within a Bayesian framework have been proposed~\cite{Smith:2017vfk, Biscoveanu:2020gds, Smith:2020lkj}. However, the work of Ref.~\cite{Renzini:2024hiu} suggests that care should be taken when using these methods, as they may be probing significantly lower redshifts compared to the standard cross-correlation searches. %This problem can be avoided by simultaneously inferring population parameters~\cite{Smith:2020lkj}, although the computational cost can be large.
Techniques that leverage principal-component analysis~\cite{Pieroni:2020rob} or neural ratio estimation~\cite{Alvey:2023npw} have been proposed in the context of LISA, and they could find application also in the context of \ac{XG} detectors.
%\Bei{I this paragraph could be moved to the end.}

%To simultaneously estimate the contributions to $\Omega_\mathrm{GW}^\mathrm{After \ Notching}(f)$ from multiple components requires us to have a deeper understanding of the unresolvable \ac{CBC} foreground.

%Since we have shown that the real limitation of notching procedure is from unresolvable \ac{CBC} foreground, and this foreground could not be further cleaned up. The technique to separate different components from the remaining data after notching is required. To simultaneously estimate the contributions to $\Omega_\mathrm{GW}^\mathrm{After \ Notching}(f)$ from multiple components requires us to have a deeper understanding of the unresolvable \ac{CBC} foreground.

%%%%%%%%%%%%%%%%%%%%%%%%%%%%%%%%%%%
%%%%%%       SECTION        %%%%%%%
%%%%%%%%%%%%%%%%%%%%%%%%%%%%%%%%%%%

\section*{Acknowledgments}
We thank Arianna Renzini for providing useful comments and suggestions on a first draft of this manuscript. The authors are grateful for computational resources provided by the LIGO Laboratory and supported by National Science Foundation (NSF) Grants PHY-0757058 and PHY-0823459. E.~Berti and L.~Reali are supported by NSF Grants No. AST-2307146, PHY-2207502, PHY-090003 and PHY-20043, by NASA Grants No. 20-LPS20-0011 and 21-ATP21-0010, by the John Templeton Foundation Grant 62840, by the Simons Foundation, and by the Italian Ministry of Foreign Affairs and International Cooperation grant No.~PGR01167. 
B.~Zhou is supported by the Fermi Research Alliance, LLC, acting under Contract No.\ DE-AC02-07CH11359.
H.~Zhong and V. Mandic were in part supported by the NSF grants PHY-2110238 and NRT-1922512.
This work was carried out at the Advanced Research Computing at Hopkins (ARCH) core facility (\url{rockfish.jhu.edu}), which is supported by the NSF Grant No.~OAC-1920103.
The authors acknowledge the Texas Advanced Computing Center (TACC) at The University of Texas at Austin for providing {HPC, visualization, database, or grid} resources that have contributed to the research results reported within this paper \cite{10.1145/3311790.3396656}. URL: \url{http://www.tacc.utexas.edu}.
%\onecolumngrid
\appendix
\section{Validity of the power-law approximation}\label{app_powerlaw}

In Fig.~\ref{fig:NC&NCS}, the attentive reader might have noticed that the foreground before notching shows a deviation from the $f^{2/3}$ power law expected for \acp{SGWB} from inspiral-dominated \acp{CBC}~\cite{Regimbau:2011bm, KAGRA:2021kbb}.
%~\Luca{What is the original paper that showed this? Regimbau 2011 is what LIGO cites...}. \eb{I think the first is Farmer-Phinney.}
This power-law approximation stems from the fact that, at leading order, the \ac{GW} amplitude for a signal in the inspiral phase scales as $A(f)\sim f^{-7/6}$~\cite{Peters:1963ux, Farmer:2003pa}. Given that the cross-correlation $C(f)$ (or equivalently the energy density $\OGW$) is proportional to $C(f)\sim f^3 |A(f)|^2$ (see Eq.~\eqref{eq:stoch}), an isotropic \ac{SGWB} generated by the superposition of inspiralling \acp{CBC} should scale as $C(f)\sim\OGW(f)\sim f^{2/3}$. 

However, waveform models such as the \texttt{IMRPhenomXAS} model considered here include higher-order corrections to the inspiral~\cite{Pratten:2020fqn}, which introduce additional power-law scalings to the expansion for the amplitude, and thus a deviation from the $f^{2/3}$ scaling. These corrections become more important close to merger and for high-mass systems.
%we choose for \ac{CBC} injections contains not only the inspiral phase but also merger and ringdown phases. The traditional understanding of $\Omega_\mathrm{GW}(f)\sim f^{2/3}$ only considers the contribution to $\Omega_\mathrm{GW}(f)$ from inspiral part of signals. However, the incorporation of merger and ringdown could result in this deviation. 
%We conduct a simple test to check this hypothesis. 
This expectation can be confirmed by injecting \ac{BBH} and \ac{BNS} events separately into two sets of time series, and computing $\hat{C}(f)$ for both sets. We show the results in Fig.~\ref{fig:simple_test}, where the left panel corresponds to \acp{BBH} and the right panel to \acp{BNS}. In both cases, we overplot the $\sim f^{2/3}$ scaling for comparison. For \acp{BBH} there is an obvious deviation from the $f^{2/3}$ scaling at frequencies $f\gtrsim 20~\rm{Hz}$, while we recover the standard behavior at lower frequencies. On the contrary, $\hat{C}(f)$ for \acp{BNS} is in good agreement with the $f^{2/3}$ power-law across the entire frequency band of interest. These results are consistent with the expectation that corrections to the leading-order scaling of the amplitude are relevant for massive systems in the later phases of their inspiral.

The $f^{2/3}$ approximation is widely used in the literature (see e.g.~\cite{KAGRA:2021kbb}), and it should be taken with care. Using this scaling, or even approximating the \ac{CBC} foreground with a single power law, may not always be a good assumption for ground-based detectors, particularly within the frequency range of current interferometers ($f\gtrsim 20-25~\rm{Hz}$). For \acp{BBH}, these scalings should be tested in greater detail using different waveform models. We leave these studies to future work.

\section{Notching results for noise-free frames}\label{app_tab}
Table~\ref{table:best_masks} shows the results for the idealized case where we inject the resolvable \ac{CBC} events with their true parameters. %into time series. 
Each row details the results for a certain baseline, with the last row showing the results combining all of the ten baselines. The second to fourth columns represent the values of $\hat{C}_\mathrm{IJ},\hat{\sigma}_\mathrm{IJ}$ and \ac{SNR} before applying any masks. The subsequent columns, from fifth to seventh, represent the results after applying the ideal masks. Table~\ref{table:biasedmasks} is similar, but we now inject resolvable \ac{CBC} events with resampled parameters. 
\begin{figure}[b]
    \centering
    \includegraphics[width=0.49\textwidth]{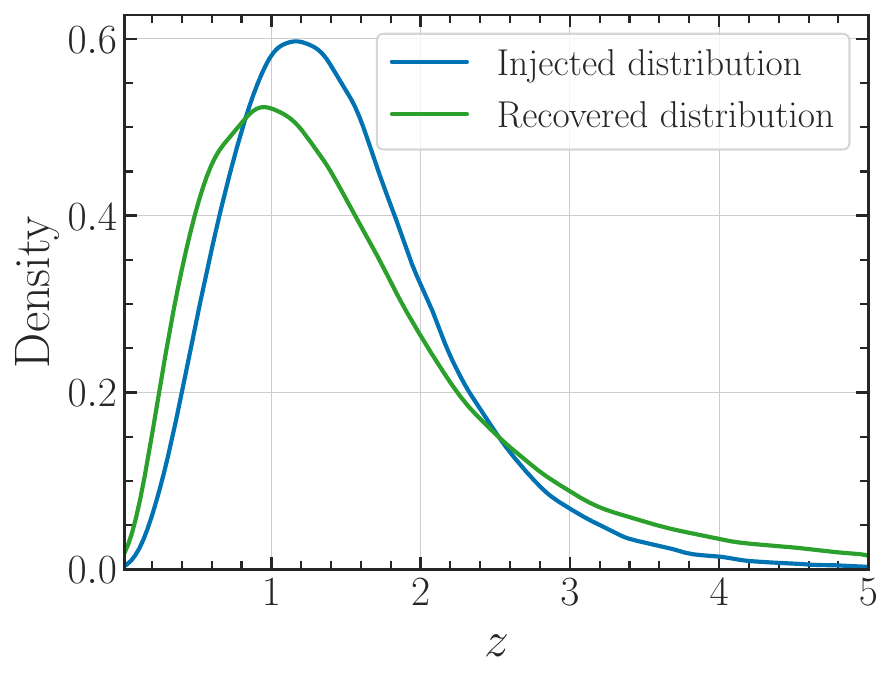}
    \caption{The injected (blue) and recovered (green) redshift distribution up to $z=5$.
      % The recovered redshifts drawn from a multivariate Gaussian distribution spread out more than the original distribution, and have a lower peak. As a result, there are more nearby (low-redshift) events in the recovered \ac{CBC} catalog than in the injected catalog.
      %\eb{I removed the last two sentences for two reasons: now the figure is in a better location, and all of this is already stated in the main text.}
     }
    \label{fig:z}
\end{figure}
\begin{table*}[!htbp]
\centering
{\footnotesize
\begin{ruledtabular}
\begin{tabular}{
  l
  S[table-format=1.1e-2]
  S[table-format=1.1e-2]
  S[table-format=1.1e-2]
  S[table-format=1.1e-2]
  S[table-format=1.1e-2]
  S[table-format=1.1e-2]
}
\toprule
\textbf{Baseline} & \multicolumn{3}{c}{\textbf{Before notching}} & \multicolumn{3}{c}{\textbf{Ideal masks applied}}\\
\cmidrule(lr){2-4} \cmidrule(lr){5-7}
& {$\hat{C}_{\rm IJ}$} & {$\hat{\sigma}_{\rm IJ}$} & {SNR} & {$\hat{C}_{\rm IJ}$} & {$\hat{\sigma}_{\rm IJ}$} & {SNR}\\
\midrule
H-L & 3.0e-10 & 1.4e-13 & 2.2e3 & 6.7e-14 & 1.6e-13 & 4.3e-1  \\
H-ET1 & 2.8e-10 & 2.2e-12 & 1.3e2 & 1.4e-13 & 2.4e-12 & 5.6e-2 \\
H-ET2 & 2.1e-10 & 1.2e-12 & 1.8e2 & 3.0e-13 & 1.8e-12 & 1.6e-1  \\
H-ET3 & 2.5e-10 & 1.3e-12 & 2.0e2 & 1.7e-13 & 1.7e-12 & 1.0e-1  \\
L-ET1 & 2.2e-10 & 1.5e-12 & 1.4e2 & 3.1e-13 & 2.3e-12 & 1.3e-1  \\
L-ET2 & 2.5e-10 & 1.0e-12 & 2.4e2 & 2.3e-13 & 1.3e-12 & 1.8e-1  \\
L-ET3 & 3.7e-10 & 2.2e-12 & 1.6e2 & 1.6e-13 & 2.4e-12 & 6.7e-1\\
ET1-ET2 & 2.8e-10 & 1.0e-12 & 2.7e2 & 7.7e-13 & 1.7e-12 & 4.5e-1  \\
ET1-ET3 & 2.9e-10 & 1.1e-12 & 2.7e2 & 7.9e-13 & 1.7e-12 & 4.5e-1  \\
ET2-ET3 & 3.0e-10 & 1.1e-12 & 2.8e2 & 8.2e-13 & 1.8e-12 & 4.6e-1 \\
\midrule
Combined &3.0e-10&1.3e-13&2.3e3&8.9e-14&1.5e-13&5.9e-1\\
\bottomrule
\end{tabular}
\end{ruledtabular}
}
\caption{Notching results for the noise-free case, where the data for cross-correlation only contains the resolvable \ac{CBC} events. The power law index $\alpha$ is set to $\alpha=0$. The first column lists baselines we consider in the analysis: H for Hanford, L for Livingston, and ET$i$ ($i=1,2,3$) for the $i$th ET detector. The second, third, and fourth columns list $\hat{C}_\mathrm{IJ}$, $\hat{\sigma}_\mathrm{IJ}$, and SNR$=\hat{C}_\mathrm{IJ}/\hat{\sigma}_\mathrm{IJ}$ before notching. The fifth, sixth, and seventh columns are the results after applying ideal masks, which are determined by the true \ac{CBC} parameters.}
\label{table:best_masks}
\end{table*}

\begin{table*}[!htbp]
\centering
{\footnotesize
\begin{ruledtabular}
\begin{tabular}{
  l
  S[table-format=1.1e-2]
  S[table-format=1.1e-2]
  S[table-format=1.1e-2]
  S[table-format=1.1e-2]
  S[table-format=1.1e-2]
  S[table-format=1.1e-2]
}
\toprule
\textbf{Baseline} & \multicolumn{3}{c}{\textbf{Before notching}} & \multicolumn{3}{c}{\textbf{Realistic masks applied}}\\
\cmidrule(lr){2-4} \cmidrule(lr){5-7}
& {$\hat{C}_{\rm IJ}$} & {$\hat{\sigma}_{\rm IJ}$} & {SNR} & {$\hat{C}_{\rm IJ}$} & {$\hat{\sigma}_{\rm IJ}$} & {SNR}\\
\midrule
H-L & 8.0e-10 & 1.4e-13 & 5.9e3 & 7.0e-14 & 1.5e-13 & 4.5e-1  \\
H-ET1 & 1.4e-10 & 2.2e-12 & 6.5e1 & 1.4e-13 & 2.4e-12 & 5.9e-2 \\
H-ET2 & 8.5e-10 & 1.2e-12 & 7.2e2 & 3.3e-13 & 1.8e-12 & 1.9e-1  \\
H-ET3 & 9.6e-10 & 1.3e-12 & 7.6e2 & 2.0e-13 & 1.7e-12 & 1.2e-1  \\
L-ET1 & 7.1e-10 & 1.5e-12 & 4.7e2 & 3.3e-13 & 2.3e-12 & 1.5e-1  \\
L-ET2 & 1.0e-9 & 1.0e-12 & 1.0e3 & 2.6e-13 & 1.3e-12 & 2.0e-1  \\
L-ET3 & 8.4e-10 & 2.2e-12 & 3.8e2 & 1.7e-13 & 2.4e-12 & 7.0e-2\\
ET1-ET2 & 7.4e-10 & 1.0e-12 & 7.2e2 & 8.3e-13 & 1.7e-12 &5.0e-1  \\
ET1-ET3 & 7.7e-10 & 1.1e-12 & 7.2e2 & 8.2e-13 & 1.7e-12 & 4.8e-1  \\
ET2-ET3 & 7.9e-10 & 1.1e-12 & 7.4e2 & 8.5e-13 & 1.7e-12 & 4.9e-1 \\
\midrule
Combined &8.0e-10&1.3e-13&6.2e3&9.5e-14&1.5e-13&6.3e-1\\
\bottomrule
\end{tabular}
\end{ruledtabular}
}
\caption{Same as Table~\ref{table:best_masks} but with realistic masks, i.e., the resolvable \ac{CBC} events are injected using the recovered parameters.}
\label{table:biasedmasks}
\end{table*}
From the second and fifth columns of Tables~\ref{table:best_masks} and~\ref{table:biasedmasks} we observe that the values of $\hat{C}_\mathrm{IJ}$ for different baselines before notching are different even when we take $\hat{\sigma}_\mathrm{IJ}$ into account. This is because the expected power-law index $\alpha$ for the \ac{SGWB} generated by an astrophysical foreground should be close to $2/3$ (possibly with some deviations, as discussed in Appendix~\ref{app_powerlaw}) while here for simplicity we use $\alpha=0$, i.e., we assume the cosmological background found after notching to be flat. This simple assumption is good enough to demonstrate that the masking procedure is efficient at removing the astrophysical foreground.

By comparing Tables~\ref{table:best_masks} and~\ref{table:biasedmasks}, we notice that the injection of resolvable \ac{CBC} events with recovered parameters leads to a much louder foreground: before notching we have $\hat{C}=3.0\times 10^{-10}$ combined among all baselines for the injection with true parameters (Table~\ref{table:best_masks}), and $\hat{C}=8.0\times 10^{-10}$ for the injection with recovered parameters (Table~\ref{table:biasedmasks}).

We explain why this is the case in Fig.~\ref{fig:z}, where we compare the injected and recovered redshift distributions for the resolved \ac{CBC} signals. We show the distribution only up to $z=5$ because the main contribution to the astrophysical stochastic foreground comes from this redshift range~\cite{Renzini:2024hiu}.
By drawing resampled parameters according to the Fisher errors, 
the resulting redshift distribution broadens, leading to a less peaked profile. Consequently, the recovered \ac{CBC} event catalog includes a larger number of nearby events compared to the original injected catalog, resulting in a louder astrophysical foreground.   

\newpage
\bibliography{references}

\end{document}